\documentclass[preprint,showpacs,preprintnumbers,amsmath,amssymb, superscriptaddress]{revtex4}

\usepackage{graphicx}
\usepackage{dcolumn}
\usepackage{bm}
\usepackage{epstopdf}
\usepackage{color}

\begin{document}

\preprint{APS/123-QED}

\title{Linear transport in stochastic media. Application to neutral transport in turbulent plasmas}

\author{Y.Marandet}
\affiliation{PIIM, CNRS/Universit\'e de Provence, Centre de Saint J\'er\^ome, Marseille F-13397 Cedex 20, France}%
 \email{yannick.marandet@univ-provence.fr}
\author{A.Mekkaoui}%
\affiliation{PIIM, CNRS/Universit\'e de Provence, Centre de Saint J\'er\^ome, Marseille F-13397 Cedex 20, France}
\author{D. Reiter}
\affiliation{IEF-4 Plasmaphysik, Forschungszentrum J\"ulich GmbH, TEC Euratom association, D-52425 J\"ulich, Germany}
\author{P. Boerner}
\affiliation{IEF-4 Plasmaphysik, Forschungszentrum J\"ulich GmbH, TEC Euratom association, D-52425 J\"ulich, Germany}
\author{P. Genesio}%
\affiliation{PIIM, CNRS/Universit\'e de Provence, Centre de Saint J\'er\^ome, Marseille F-13397 Cedex 20, France}
\author{J. Rosato}
\affiliation{PIIM, CNRS/Universit\'e de Provence, Centre de Saint J\'er\^ome, Marseille F-13397 Cedex 20, France}
\author{H. Capes}
\affiliation{PIIM, CNRS/Universit\'e de Provence, Centre de Saint J\'er\^ome, Marseille F-13397 Cedex 20, France}
\author{F. Catoire}
\affiliation{PIIM, CNRS/Universit\'e de Provence, Centre de Saint J\'er\^ome, Marseille F-13397 Cedex 20, France}
\author{M. Koubiti}
\affiliation{PIIM, CNRS/Universit\'e de Provence, Centre de Saint J\'er\^ome, Marseille F-13397 Cedex 20, France}
\author{L. Godbert-Mouret}
\affiliation{PIIM, CNRS/Universit\'e de Provence, Centre de Saint J\'er\^ome, Marseille F-13397 Cedex 20, France}
\author{R. Stamm}
\affiliation{PIIM, CNRS/Universit\'e de Provence, Centre de Saint J\'er\^ome, Marseille F-13397 Cedex 20, France}

\date{\today}

\begin{abstract}
This work addresses linear transport in turbulent media, with emphasis on neutral particle (atoms, molecules) transport in magnetized fusion plasmas. A stochastic model for turbulent plasmas, based upon a multivariate Gamma distribution, is presented. The geometry is a 2D slab and turbulence is assumed to be statistically homogeneous. The average neutral density and  ionization source, which are the quantities relevant for integrated simulations and diagnostic applications, are calculated analytically in the scattering free case. The boundary conditions and the ratio of the turbulence correlation length to the neutral mean free path are identified as the main control parameters in the problem. The non trivial relationship between the average neutral density and the ionization source is investigated. Monte Carlo calculations including scattering are then presented, and the main trends obtained in the scattering free case are shown to be conserved.  
\end{abstract}

\maketitle

\section{Introduction}

The problem of linear transport in a stochastic background medium has attracted much interest in the last twenty years, with applications ranging from neutronics in boiling fluids to radiation transport in cloudy atmospheres \cite{Pomraning91}.  Technically, the problem consists in treating the absorption and scattering rates, which describe the interaction of particles with the host medium in the linear transport equation, as random variables. The resulting stochastic integro-differential equation is then solved for the statistical moments of the particle distribution function. The back-reaction of transport on the host medium is neglected, and the statistical properties of the latter have to be specified for each particular problem. The most thoroughly studied case is a mixture of two fluids, where the rates are discrete two states random variables \cite{Pomraning91}.  The case of Gaussian statistics has subsequently been studied, as a first attempt to address the problem of neutral particle transport in fusion plasmas \cite{Prinja93,Prinja96}, where plasma turbulence leads to a continuous probability density function for the rates. In fact, in magnetic fusion devices such as tokamaks, the transport of species such as atoms or molecules, as well as line radiation, is best described in kinetic terms. The single particle distribution function $f(\mathbf{v},\mathbf{r},t)$ obeys a Boltzmann equation, which is linear provided neutral-neutral collisions are negligible. In practice, given the complexity of the geometry and the numerous reaction channels involved, Boltzmann's equation is often solved by a Monte Carlo approach \cite{Spanier08}. Several computer codes have been developed also in the magnetic fusion context, among which the EIRENE code \cite{Reiter05} which is used for ITER modeling, often iteratively in conjunction with the B2 plasma fluid code \cite{Braams86}. Indeed, at the outer plasma region (the so called ``plasma edge''), where the density of neutrals can be of the same order as the electron density, the fluid equations describing the plasma and the kinetic neutral transport problem must be solved consistently. These models have a mesoscopic character, in the sense that the time scales of interest (of the order of 10 ms) are large compared to time scales characterizing turbulence (correlation time of about 10 $\mu s$ \cite{Boedo09}), and small compared to macroscopic times in the discharge (\textit{e.g.} seconds). The effect of turbulence at these mesoscopic scales is
essential, since turbulent transport usually dominates collisional transport \cite{Bourdelle05}. Turbulent transport is described in current fluid codes by \textit{ad-hoc} empirical transport coefficients $(D_{\perp},v_{\perp})$ in the direction transverse to the magnetic field, such that for the particle flux $\Gamma_{turb}=\langle \tilde{N}\tilde{v}\rangle=-D_{\perp}\nabla \langle N\rangle+v_{\perp}\langle N\rangle$, where brackets denote a time average ($\tilde{N}$ and $\tilde{v}$ stand for the fluctuating part of the density and the velocity fields). However, neutral transport is still commonly calculated on the average plasma background, without any account of the underlying turbulence. The problem is especially acute at the plasma edge, since edge plasma turbulence has different properties compared to core turbulence (prevailing in the central part of magnetically confined plasmas). In particular, fluctuation rates up to order unity are observed in the outer edge region of the plasma, where magnetic field lines are ``open'' (scrape off layer, SOL), \textit{i.e.} intersect solid material components. These large fluctuation rates are often interpreted in terms of propagating plasma filaments (aka ``blobs'' or ``avalanches''), which have been seen both experimentally \cite{Terry03,Grulke06} and numerically \cite{Sarazin03}. In the drift plane, perpendicular to the magnetic field, these filaments have an extension of about 1 cm. They propagate radially outward with a velocity of the order of 1 km/s \cite{Boedo09}. The question of the existence of a significant contribution of turbulence to neutral particle transport thus naturally arises. Moreover, plasma spectroscopy usually works with integration times much larger than the turbulent time scales, so that the measured signals, such as Doppler broadened spectral line profiles, are time and space averages over fluctuations \cite{Marandet09}. As a result, the observed Doppler profile is related to the average kinetic velocity distribution $\langle f\rangle$. In the following, the time averages involved in either transport or spectroscopic data modeling will be replaced by an ensemble average over a number of realizations of the turbulent density fields. The most attractive feature of such a stochastic model is its flexibility, which allows to fully explore the physics of the problem, and in particular to study limiting cases. However, the model parameters can also be specified from experimental and/or numerical results. The outline of the work is the following. 
Section \ref{sec:1} is devoted to setting up the problem of kinetic transport in a stochastic background. In section \ref{sec:2}, the stochastic model used to describe the turbulent background medium is presented. This model, which relies on a multivariate Gamma distribution, is consistent with experimental data and alleviates difficulties encountered with Gaussian statistics. Section \ref{sec:3} deals with the numerical implementation of the model in the EIRENE Monte Carlo solver, and its validation against analytical results. Finally numerical calculations including scattering are presented.

\section{\label{sec:1}Stochastic linear Boltzmann equation}

\subsection{General formulation}

The transport of neutral particles in a
fusion plasma is best described in
terms of a Boltzmann equation

\begin{equation}\label{eq:Boltzmann}
\left(\frac{\partial}{\partial t}+v\boldsymbol{\Omega}\cdot\nabla\right) f(\mathbf{r},v,\boldsymbol{\Omega},t)=\left(\frac{\partial f}{\partial t}\right)_c,
\end{equation}

\noindent where $f(\mathbf{r},v,\boldsymbol{\Omega},t)$ is the
neutral particles velocity  distribution at point $\mathbf{r}$,
$\mathbf{v}$ the velocity and $\boldsymbol{\Omega}$ the
unit vector along $\mathbf{v}$, such that $\mathbf{v}=v\boldsymbol{\Omega}$. Inhomogeneous terms due
to primary sources of neutral particles, are omitted here and in the
following discussions, for simplicity. For ionizing conditions, the collision
operator on the r.h.s. of the previous equation is given for
hydrogen isotopes atoms by

\begin{equation}\label{eq:collision_term}
\left(\frac{\partial f}{\partial t}\right)_c=-(\nu_{io}+\nu_{cx})f+\int_0^{+\infty}v'^2 dv' \int d\boldsymbol{\Omega}' \sigma_{cx}(|\mathbf{v}-\mathbf{v}'|) |\mathbf{v}-\mathbf{v}'| f(\mathbf{r},v',\boldsymbol{\Omega}')  f_i(\mathbf{r},v,\boldsymbol{\Omega}),
\end{equation}

\noindent where $\nu_{io}$ and $\nu_{cx}$ are respectively the
ionization (by electron impact) and charge-exchange rates, $\sigma_{cx}$
is the charge exchange cross section, and $f_i$
the ion velocity distribution. The rates are defined by
$\nu_{io}=N_e\overline{\sigma_{io} v_e}$ and
$\nu_{cx}=N_i\overline{\sigma_{cx} |v-v_i|}$ where the overbars
respectively stand for an average over the electron and the ion
velocity distributions (assumed to be maxwellian), and $\sigma_{io}$ is the ionization cross section. For $H_2$  (and $D_2$, $DH$, ...) molecules, a reasonable first approximation is

\begin{equation}\label{eq:collision_term_mol}
\left(\frac{\partial f_{mol}}{\partial t}\right)_c=-\nu_{d}f_{mol},
\end{equation}

\noindent where $\nu_d$ is an effective dissociation rate, including contributions from ionization to $H_2^+$, dissociation by electron impact, dissociative ionization, possibly ion conversion (see \cite{Kotov06} for an overview and more advanced models). A source term describing atoms formed by molecular dissociative processes should be added to Eq. (\ref{eq:collision_term})   to treat both species simultaneously, but this will not be addressed analytically. Eq. (\ref{eq:Boltzmann}) together with Eq. (\ref{eq:collision_term}) (or Eq. (\ref{eq:collision_term_mol})) form the basis of the linear transport problem \cite{Duderstadt79,CaseZweifel67} (linear in the sense that the collision term given by Eq. (\ref{eq:collision_term}) and (\ref{eq:collision_term_mol}) are linear in $f$). In a general theoretical setting, ionization and charge exchange would respectively be called absorption and scattering. Although in the following we discuss neutral particle transport in turbulent plasmas, most of the results are presented  in normalized units, so as to be independent on the precise value of the rates.  Therefore, these results could be of general interest for linear transport theory.

The rate coefficients and rates (frequencies) are known functions of the
plasma density and temperature, and
in case of charge exchange also of the neutral velocity $v$. When the plasma is
turbulent, the values of the rates
also reflect the plasma fluctuations,
thus introducing a coupling between plasma turbulence and neutral particle
transport. In this work, we will neglect the back reaction of the resulting
fluctuations of the neutral density or temperature on the plasma turbulence
itself, \textit{i.e.} neutral particles will be treated as a passive species. The statistical
properties of turbulence are then input quantities, determined by independent
turbulence modeling or inferred from experiments. In the following, we will assume that the problem is stationary in time, that is  the time derivative will be neglected in Eq. (\ref{eq:Boltzmann}). This
approximation is justified whenever the typical life time of a
neutral particle, given by $\nu_{io}^{-1}$, is much smaller than
the typical time over which turbulent fields evolve  (a few
microseconds). This is in general not the case in fusion edge plasmas, but this
approximation is very useful to understand the physics. It could be relaxed in the numerical approach presented in
section \ref{sec:3}, which relies on the Monte Carlo code EIRENE. For the time being, we are interested in the time average of the neutral distribution over durations much larger than the turbulence correlation
time, so that according to the ergodic theorem, time averages can be
replaced by ensemble averages provided the statistics is stationary
\cite{Lumley70}. Averaging Eq. (\ref{eq:Boltzmann}) over turbulent
fluctuations yields

\begin{equation}\label{eq:Boltzmann_av}
v\boldsymbol{\Omega}\cdot\boldsymbol{\nabla}\langle f\rangle=-\left\langle(\nu_{io}+\nu_{cx})f\right\rangle+\int_0^{+\infty}v'^2 dv' \int d\boldsymbol{\Omega}'\sigma_{cx}(|\mathbf{v}-\mathbf{v}'|) |\mathbf{v}-\mathbf{v}'| \left\langle f(\mathbf{r},v',\boldsymbol{\Omega}')  f_i(\mathbf{r},v,\boldsymbol{\Omega})\right\rangle.
\end{equation}

\noindent This is clearly not a
closed equation for $\langle f\rangle$, since the averages on the
r.h.s are not given as functions of  $\langle f\rangle$. Neglecting
turbulence, as is currently common practice in neutral particle transport modeling for tokamaks (\textit{e.g.} in all B2-EIRENE, EDGE-2D-EIRENE or EMC3-EIRENE applications) for neutrals, amounts to solve Boltzmann's equation using the value of the rates calculated for the average values of the turbulent fields. For instance, the first term on the r.h.s. of Eq. (\ref{eq:Boltzmann_av}) is in fact currently approximated by

\begin{equation}
\left\langle(\nu_{io}+\nu_{cx})f\right\rangle \simeq
\left[\nu_{io}(\langle N_e\rangle,\langle T_e\rangle)+\nu_{cx}(\langle
N_i\rangle,\langle T_i\rangle)\right]\langle f\rangle. 
\end{equation}

\begin{figure}[htb] 
\includegraphics[width=.3\textwidth]{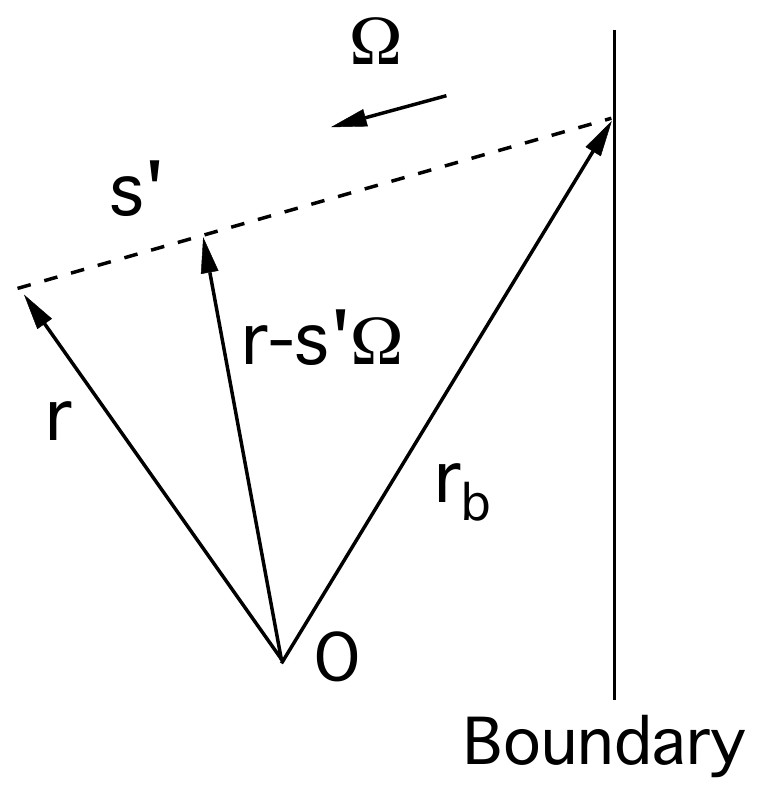}
\caption{Sketch of the different vectors involved in the integral formulation of the problem. $\mathbf{r}$ defines the position at which the neutral particle distribution is evaluated, ${\mathbf{r_b}}$ the position of the boundary along the neutral flight assuming their velocity is parallel to the unit vector $\boldsymbol{\Omega}$. s' measures the distance of the neutral flight, starting from $\mathbf{r}$, \textit{i.e.} going backwards along the neutral particle trajectory. \label{fig:vectors}}
\end{figure}

\noindent Assuming stationarity, Eq. (\ref{eq:Boltzmann}) is equivalent to the following integral equation 

\begin{equation}\label{eq:Boltzmann_int}
f(\mathbf{r},v,\boldsymbol{\Omega})= f_0 e^{-\tau(s)} + \int_0^{s} \frac{ds'}{v}  Q_{cx}(\mathbf{r}-s'\boldsymbol{\Omega},v,\boldsymbol{\Omega)}e^{-\tau(s')},
\end{equation}

\noindent where $f_0=f(\mathbf{0},v,\boldsymbol{\Omega})$, and $Q_{cx}$ is the source of neutrals with velocity $v$ in direction $\boldsymbol{\Omega}$ at point $\mathbf{r}$ due to charge exchange, given by

\begin{equation}
Q_{cx}(\mathbf{r},v,\boldsymbol{\Omega})=\left[\int_0^{+\infty}v'^2dv' \int d\boldsymbol{\Omega}' \sigma_{cx}(|\mathbf{v}-\mathbf{v}'|) |\mathbf{v}-\mathbf{v}'| f(\mathbf{r},v',\boldsymbol{\Omega}')\right]  f_i(\mathbf{r},v,\boldsymbol{\Omega}),
\end{equation}

\noindent and $s$ the distance to the boundary along direction $\boldsymbol{\Omega}$, \textit{i.e.} $s=|\mathbf{r}-\mathbf{r}_b|$ (where $\mathbf{r}_b$ is the intersection between the neutral particle trajectory and the boundary, see Fig. \ref{fig:vectors}). $\tau(s)$ is the
optical thickness defined by

\begin{equation}\label{eq:opt_thick}
\tau(s)=\int_0^{s} \nu(\mathbf{r}-s'\boldsymbol{\Omega})\frac{ds'}{v},
\end{equation}

\noindent where $\nu=\nu_{io}+\nu_{cx}$ is the total
attenuation rate. 
The exponential terms in Eq. (\ref{eq:Boltzmann_int}) give the
probability that a neutral particle does not undergo a collision during a
flight at least of length $s$, so that Eq. (\ref{eq:Boltzmann_int}) has a natural probabilistic interpretation. Finally, Eq. (\ref{eq:Boltzmann_av}) is equivalent to the following integral
equation

\begin{equation}\label{eq:Boltzmann_av_int}
\langle f(\mathbf{r},v,\boldsymbol{\Omega})\rangle=\left\langle f_0 e^{-\tau(s)} \right\rangle + \int_0^s \frac{ds'}{v} \left\langle Q_{cx}(\mathbf{r}-s'\boldsymbol{\Omega},v,\boldsymbol{\Omega)}e^{-\tau(s')}\right\rangle,
\end{equation}

\noindent obtained by averaging Eq. (\ref{eq:Boltzmann_int}) over the background fluctuations. The average neutral density is defined by

\begin{equation}
\langle N(\mathbf{r})\rangle =\int_0^{\infty} v^2 dv \int_{4\pi}d\boldsymbol{\Omega} \ \langle f(\mathbf{r},v,\boldsymbol{\Omega})\rangle.
\end{equation}

\noindent In the case of atoms in a plasma, $\langle N(\mathbf{r})\rangle$ may be determined by Laser Induced Fluorescence (LIF) \cite{Mertens97},  by averaging successive measurements performed in a plasma in statistical steady state. The primary quantity of interest for integrated edge plasma engineering codes is the average ionization source $\langle S(\mathbf{r})\rangle=\langle \nu(\mathbf{r}) N(\mathbf{r})\rangle$, which provides the local particle source in the plasma equations. Its volume integral is equal to the average of the source flux, thus ensuring that the total neutral content $\langle N_{tot}\rangle$ is stationary. It should be emphasized that $\langle S\rangle$ is in general NOT related to the average density in a simple and obvious way, in other words: $\langle S\rangle \ne \langle\nu\rangle \langle N\rangle$. Instead, $\langle N_{tot}\rangle$ depends on the properties of the fluctuations. In the context of radiation transport, $\langle S\rangle$ would for instance be an average photo-excitation rate.
\subsection{Geometry}

We consider a 2D slab geometry, of size $\mathcal{L}\times \mathcal{L}$ in the $xy$ plane. The particle source is located at $x=0$, and will be specified in section \ref{sec:boundary}. This geometry will be further degraded to a 1D problem as a first step in the analytical calculations. When dealing with neutral particle transport in tokamak plasmas in particular, the $x$ axis
represents the radial direction, along the minor radius of the torus. The $y$ axis is the poloidal direction, and the $z$ direction is along the magnetic field (ignoring the typically small field line pitch). Turbulent wave numbers in the parallel direction $k_{\parallel}$ are such that $k_{\parallel}\gg k_{\perp}$, so that the plasma is taken to be homogeneous along $z$ ("flute approximation"), which justifies a 2D model \cite{Sarazin03}. 3D effects are believed to be essential to reproduce experimental features of edge turbulence \cite{Tamain09}, but for our purposes where the statistical properties of turbulence are imposed, the 2D approximation is sufficient. In the following, our work will focus on the average neutral density and ionization source as a function of the radial coordinate $x$. $\langle N (x)\rangle$ is related to the flux of neutrals crossing a given radial surface $x=cst$. In the simplest case, \textit{i.e.} one speed scattering free transport problem in 1D, $\Gamma(x)=\langle N(x)\rangle v_0$ where $v_0$ is the neutral velocity. As a result, the screening efficiency $p(x)$ of slab of thickness $x$, $p(x)=1-\Gamma(x)/\Gamma(0)$, is directly linked to $\langle N (x)\rangle$. 

\subsection{Stochastic model for turbulent plasma background\label{subsec:stochastic_model}}

The statistical properties of the rate
coefficients are  directly related to those of the plasma
parameters, since for example $\nu=\nu_{io}(N_e,T_e)$. In principle, calculating the averages in Eq. (\ref{eq:Boltzmann_av_int}) requires knowledge of the functional PDF for the field $\nu(\mathbf{r})$, $W[\nu]$. In practice, this functional is not known, but the PDF at one point $P_1(\nu,\mathbf{r})$ ($P_1(\nu,\mathbf{r})d\nu$ gives the probability that the rate takes a value between $\nu$ and $\nu+d\nu$ at point $\mathbf{r}$ ), and the 2 point correlation
function $C(\mathbf{r},\mathbf{r}')=\langle\nu(\mathbf{r})\nu(\mathbf{r}')\rangle - \langle\nu(\mathbf{r})\rangle\langle\nu(\mathbf{r}')\rangle$ (related to the spatial
spectra obtained by a Fourier transform) can be measured or calculated numerically. In the following, we assume statistical homogeneity and isotropy, so that $C(\mathbf{r},\mathbf{r}')=C(|\mathbf{r}-\mathbf{r}'|)$. To avoid unnecessary complications related to functional integration, we discretize the problem in space, and focus on a 1D geometry first. The number of cells $n$ in the spatial grid has to be chosen large enough so that the cell size $\epsilon$ is much smaller than the turbulence correlation length $\lambda$. In this approximation, $W[\nu]$ becomes a multivariate distribution $W_n(\boldsymbol{\nu})$ where $\boldsymbol{\nu}=(\nu_1,\ldots,\nu_n)$, $\nu_i$ being the rate value in cell number $i$. The simplest choice for $W_n$ would seem to be a Gaussian multivariate distribution, as in references \cite{Prinja93,Prinja96}. This has the advantage to permit an exact analytical closure of Eq. (\ref{eq:Boltzmann_av}), through an elegant use of Novikov's theorem \cite{Novikov65}. The average velocity distribution $\langle f\rangle$   then obeys an effective Bolzmann equation, in which the absorption and scattering rates are renormalized by fluctuations. However, the absorption rate becomes negative for large fluctuation rates, a consequence of an unphysical feature of the model. In fact, for Gaussian statistics, there are realizations for which the rates are negative, at least in some regions of space. The weight of these negative values is all the larger that the fluctuation rate increases. This problem is very well illustrated in Ref. \cite{Prinja95}, where it is shown that the Gaussian model might nevertheless provide interesting qualitative information on the physics of the problem. In summary, the main drawback of the Gaussian model is its limited ability to describe the large fluctuation rate regime, where the effects of fluctuations become really significant. Truncating the Gaussian distribution to positive variables is not a good solution, because the nice properties of Gaussian statistics are then lost. In particular, even in the scattering free case, this choice precludes analytical calculations, since the integrals to be calculated are then orthant probabilities (\cite{Kotz}, p. 127). As a result, in the following we will consider a statistical model involving only positive random variables.  Moreover, recent experimental investigations of edge turbulence have shown that density fluctuations in the SOL are not Gaussian, and suggest at least two possible choices. In the SOL, the PDF $P_1$ has been found to be well fitted by either Log-normal or Gamma distributions (\textit{e.g.} \cite{Graves05}). Taking $W_n$  as multivariate Log-normal distribution \cite{Kotz} precludes analytical calculations of the averages involved in Eq. (\ref{eq:Boltzmann_av_int}). Therefore, for validation purposes, in the following we focus on a multivariate generalization of the Gamma distribution (in fact of the chi-2 distribution), which is the marginal of the Wishart $n\times n$ matrix distribution, for which only the $n$ diagonal elements are retained \cite{Krishnaiah61}. To define the multivariate Gamma distribution, first consider the following $n$-variate Gaussian PDF

\begin{equation}\label{eq:multivariate_gauss}
P_n(\mathbf{X})=\frac{1}{(2\pi)^{n/2}\sqrt{\det(\mathbf{G})}}\exp\left(-\frac{1}{2}\mathbf{X}^T \mathbf{G}^{-1}\mathbf{X}\right),
\end{equation}

\noindent where $G_{ij}=\langle X_i X_j\rangle$ and $\langle X_i\rangle=0$. Let $X_{ij}$,
$i=1,\ldots,n$ and $j=1,\ldots,M$ be $M$ independent series of $n$ zero average
Gaussian random numbers sampled from $P_n$ (\textit{i.e.} such that $\langle X_{ij}
X_{lm}\rangle=G_{il}\delta_{jm}$),  then the $n$ variables

\begin{equation}\label{eq:Yi}
\nu_i=\sum_{j=1}^M X_{ij}^2,
\end{equation}

\noindent are distributed according to $W_n(\boldsymbol{\nu})$. The latter is such that its 1 point marginal $W_1(\nu_i)$ is a chi-2 distribution with M degrees of freedom, \textit{i.e.} a Gamma distribution of shape parameter $\beta=M/2$ (see Appendix A), and its 2  point  correlation function is given by $\langle\langle \nu_i \nu_j\rangle\rangle=\langle\nu_i\nu_j \rangle-\langle\nu_i\rangle\langle \nu_j\rangle=2MG_{ij}^2$. In fact from Eq. (\ref{eq:Yi}), we have $\langle \nu_i \rangle = M G_{ii}$ and

\begin{equation}
\langle\langle \nu_k \nu_l\rangle\rangle=\sum_{i=1}^M \langle X_{ki}^2 X_{lj}^2 \rangle + \sum_{i\neq j} \langle X_{ki}^2 \rangle \langle X_{lj}^2 \rangle-M^2 G_{kk}G_{ll}.
\end{equation}

\noindent Expressing the fourth moment $\langle X_{ki}^2 X_{lj}^2 \rangle$ in terms  of the second moments, we get $\langle\langle \nu_k\nu_l\rangle\rangle=2MG_{kl}^2$. In other words, the correlation matrix of the $\boldsymbol{\nu}$ variables is $C_{ij}=2MG_{ij}^2$, so that it is fully specified by $G_{ij}$. However, note that $G_{ij}$ is not defined in a unique way from $C_{ij}$. Moreover, $C_{ij}\ge 0$ by construction, so that the multivariate Gamma model cannot describe situations where the correlation function oscillates around 0. As an example, to obtain the correlation matrix  $\mathbf{C}$ corresponding to

\begin{equation}\label{eq:correl_function_exp}
C(\mathbf{r}-\mathbf{r}')=\sigma^2 \exp(-|\mathbf{r}-\mathbf{r}'|/\lambda),
\end{equation}

\noindent where $\sigma$ is the standard deviation, and
$\lambda$ the turbulence correlation  length (or  integral scale), one should set $C_{ij}=C(\mathbf{r}_i-\mathbf{r}_j)$, where $\mathbf{r}_i$ and
$\mathbf{r}_j$ are the coordinates of the center of the cells $i$
and $j$, respectively. Then,

\begin{equation}\label{eq:G_exp}
G_{ij}=\frac{\sigma}{\sqrt{2M}}\exp(-|\mathbf{r}_i-\mathbf{r}_j'|/ 2\lambda).
\end{equation}

\noindent The fluctuation rate $R$ (\%) is determined by $M$ through $R=\sqrt{2/M}\times100$, and in the following we shall limit ourselves to $M\geq 3$, so that $W_1(\nu_i=0)=0$. It should be noted that an integral scale can be defined for any correlation function by

\begin{equation}
\lambda=\int_0^{+\infty}C(r)dr/C(0),
\end{equation}

\noindent provided the integral converges. $\lambda$ represents the typical size of the turbulent structures. The case where $\lambda$ is infinite, because of long range correlations (\textit{e.g.} $C(r)\propto r^{-\alpha}$, $\alpha\leq 1$), will not be considered here. Instead, the limit $\lambda\rightarrow +\infty$ should be understood as a situation where the plasma background becomes spatially homogeneous, see Fig. \ref{fig:maps_ex}. The most useful form of $W_n(\boldsymbol{\nu})$ for our purposes is the following,

\begin{equation}\label{eq:multi_chi-2}
W_n(\boldsymbol{\nu})=\int\ldots\int \left[\prod_{j=1}^M d^n\mathbf{X}_j \ \ P_n(\mathbf{X}_j)\right] \prod_{l=1}^n \delta\left(\nu_l-\sum_{j=1}^M X_{lj}^2\right),
\end{equation}

\noindent where $P_n$ is given by Eq.(\ref{eq:multivariate_gauss}) . Eq. (\ref{eq:multi_chi-2}) is just the mathematical translation of the definition of the statistics of $\nu_i$. The generalization to a 2D domain consisting of $n\times n$ cells can be made in two equivalent ways, either by using Eq. (\ref{eq:multi_chi-2}) with $n$ replaced by $n^2$ and a proper labeling scheme for the cells, or by using Wishart's distribution (for which the random variable is a $n\times n$ matrix) \cite{Kotz}.

\subsection{Sources\label{sec:boundary}}

We first address the question of boundary conditions (\textit{i.e.} more precisely sources in the case where charge exchange is included) from the physics rather than the technical point of view, hence focusing on neutral particles in plasma. The main source of neutrals in a magnetic fusion confinement device is recycling, that is ions hitting the wall subsequently reenter the plasma as neutrals \cite{Stangeby00}. Recycling is a complex process, involving several different elementary mechanisms, the time scales of which are not necessarily short compared to those of turbulence. For instance, ions impinging on the wall can be backscattered as neutrals, and this fraction of the recycling flux will be directly related to the instantaneous plasma flux because time scales involved are very short ($<10^{-9}$ s). The backscattering probability is of the order of 0.1 for hydrogen on carbon material surfaces, and can be significantly larger for heavier fusion relevant materials like tungsten \cite{Stangeby00}. Hydrogen ions can, after neutralization, also thermalize with the wall, then recombine with another atom in a H$_2$ molecule,  which then desorbs. In this case, time scales involved depend on the wall properties, \textit{i.e.} whether it is saturated with hydrogen or not. In the latter case, the wall acts as a pump and its response becomes very slow \cite{Wienhold84}. As a result, in the following, we will limit ourselves to two extreme cases that will be called slow and fast recycling. In the first case, recycling is slow compared to turbulence, so that neutrals are emitted homogeneously from the wall (the average plasma flux is assumed to be homogenous along $y$). In this case, the boundary condition at the wall is a non stochastic and uniform flux at the wall. The second case corresponds to a situation where recycling is much faster than turbulence, so that the local neutral flux at the wall is a direct reflection of the plasma flux distribution $\Gamma_p$. The latter is assumed to be given by $\Gamma_p(y)=N_e(0,y) V_{blob}$ (unit m$^{-2}$.s$^{-1}$), where $V_{blob}$ is the typical "plasma blob" velocity, of the order of $V_{blob}\simeq 1$ km/s \cite{Boedo09}. In reality if the wall is not saturated, a superposition of these two limiting cases may provide a reasonable description of the recycling source. Finally, it should be pointed out that in the frame of radiation transport, the fast recycling case would describe a reflective boundary, while the slow recycling case would represent the surface of a black body.

\section{\label{sec:2} Analytical case for validation}

In order to obtain a model which can be solved analytically, we consider the scattering free (\textit{i.e.} no charge exchange) problem in a 2D slab geometry. Only density fluctuations are considered, and the ionization rate $\nu$ is assumed to be linear in density (which physically means that the multi-step effects of the collisional radiative ionization cascade are neglected). In this case, the multivariate PDFs of the rate coefficient and of the density are identical up to a scaling factor. We obtain the solution in two steps. In the scattering free problem, neutral particles move along straight lines, and we start by solving for the neutral velocity distribution along a given direction, defined by the unit vector $\boldsymbol{\Omega}$ (1D problem). We further assume that the source is  mono-kinetic, with a velocity $v=v_0$ (one speed transport problem). In this simplified case, Eq. (\ref{eq:Boltzmann_av_int}) reduces to

\begin{equation}\label{eq:VDF_simplified}
\langle f(s,v,\boldsymbol{\Omega})\rangle=\frac{1}{v_0}\delta(v-v_0)\left\langle \Gamma_0 e^{-\tau(s)}\right\rangle,
\end{equation}

\noindent  where $\tau$ is the optical thickness defined by Eq. (\ref{eq:opt_thick}). $\Gamma_0$ is the flux density (unit m$^{-2}$.s$^{-1}$) at the wall ($x=0$), which is either non-stochastic (slow recycling) or proportional to the instantaneous plasma density at the wall (fast recycling). The average neutral particle density in 2D $\langle N_{2D}(x)\rangle$ is obtained by first averaging Eq. (\ref{eq:VDF_simplified}) on $v_0$ and $\boldsymbol{\Omega}$, then integrating over $v$.  The average ionization source is given by

\begin{equation}\label{eq:Sp_simplified}
\langle S(x)\rangle=\langle \nu(x)N(x)\rangle=\frac{1}{v_0}\left\langle \Gamma_0 \nu(x) e^{-\tau(x)}\right\rangle.
\end{equation}

\noindent The scattering free case is very interesting both for code validation purposes and for understanding the physics of the problem. Moreover, it provides a reasonable approximation for neutral species sputtered from the wall, and for $H_2$ molecules for which scattering can indeed be neglected, except in high density and low temperature plasmas ($N_e>10^{20}$ m$^{-3}$ and $T_e<5$ eV), where elastic collisions have to be retained \cite{Kotov06}.

\subsection{General results}

There are two important results which are valid for any choice of the turbulence statistical properties. The first one pertains to the vanishing turbulence correlation length $\lambda$ limit, that is when the ratio $a$ of $\lambda$ to the neutral mean free path $l$ tends to zero. In fact, the quantity $\tau(s)/s$, where $\tau(s)$  is the optical thickness defined by Eq. (\ref{eq:opt_thick}), is nothing else but the 1D spatial average of $\nu(\mathbf{r}-s'\Omega)/v$ along the neutral particle trajectory.  In the limit where $\lambda/s \rightarrow 0$, the ergodic theorem \cite{Lumley70} ensures that for almost all realizations of $\nu$

\begin{equation}
\frac{1}{s}\int_0^s \nu((\mathbf{r}-s'\Omega)/v_0)ds'=\langle \tau(s)\rangle,
\end{equation}

\noindent which implies

\begin{equation}\label{eq:lambda_vanish}
\langle f(x)\rangle_{\lambda=0} = \frac{\langle\Gamma_0\rangle}{v_0}\delta(v-v_0) e^{-\langle \tau(x)\rangle}.
\end{equation}

\noindent Therefore, in the $\lambda\rightarrow 0$ limit, the average neutral distribution obeys an effective (scattering free) Boltzmann equation in which the ionization rate is simply replaced by its average value. In other words, the crossed term $\langle \nu f \rangle$ factorizes in $\langle \nu \rangle \langle f \rangle$. This limit is called the atomic mix limit in Ref. \cite{Pomraning91}, where it is obtained by different arguments. From a physical point of view, at a distance $x\gg \lambda$ from the source, the correlations between the values of $f$ and $\nu$ are expected to be negligible. Indeed, $f$ depends on the values taken by $\nu$ between $0$ and $x$. It should be noted that in general the ionization rate $\nu$ is a non linear function of the plasma parameters ($N_e$, $T_e$), so that $\langle\nu(N_e,T_e)\rangle\neq \nu(\langle N_e\rangle,\langle T_e\rangle)$. Therefore, the turbulence free case is recovered in the $\lambda\rightarrow 0$ limit only when $\nu$ is a linear function of $N_e$ (and more generally, when temperature fluctuations are neglected). Now, we come to the second important point. Jensen's inequality \cite{Gradshteyn}, which pertains to convex functions, states that

\begin{equation}\label{eq:Jensen}
\left\langle e^{-\tau(x)} \right\rangle \geqslant e^{-\langle \tau(x)\rangle}.
\end{equation}

\noindent The equality is realized in the $\lambda=0$ limit (see Eq. (\ref{eq:lambda_vanish})), \textit{i.e.} the fastest decaying average density profile is obtained for vanishing correlation lengths. In the model considered in this section (one speed transport, no scattering, $\nu$ linear in density), we have thus proved that, provided temperature fluctuations are neglected, the optical thickness of the plasma $\tau(s)$ is always reduced by fluctuations.

\subsection{Characteristic function}

The quantities we are interested in, namely the average density $\langle N\rangle$ and ionization source $\langle S\rangle=\langle \nu N\rangle$, are related to the characteristic functional associated to $W[\nu]$,

\begin{equation}
Z[u]=\left\langle\exp -\int_0^{+\infty} \nu(y)u(y)dy \right\rangle,
\end{equation}

\noindent  and to its functional derivatives with respect to $u$, where $u(y)$ is an arbitrary function of $y$. For instance,

\begin{equation}\label{eq:dZfunc}
\frac{\delta Z[u]}{\delta u(a)}(u\equiv\Theta(x'-x)/v_0)= - \left\langle \nu(a) \exp -\int_0^{x} \nu(y)\frac{dy}{v_0} \right\rangle,
\end{equation}

\noindent where $\Theta(x'-x)$ is the Heaviside step function. It turns out that the discretized version of $Z[\nu]$, $Z_n(\boldsymbol{\nu})$ with $\boldsymbol{\nu}=(\nu_1,\ldots,\nu_n)^T$ being a column vector, can be calculated analytically, provided the rate coefficients statistics is given by a multivariate Gamma distribution. $Z_n$ is defined by

\begin{equation}
Z_n(\mathbf{u})=\int d\boldsymbol{\nu} W_n(\boldsymbol{\nu})\exp -\frac{\epsilon}{v_0}\boldsymbol{\nu}\cdot\mathbf{u},
\end{equation}

\noindent where $\mathbf{u}=(u_1,\ldots,u_n)^T$ and $\epsilon$ is the discretization step. Plugging Eq. (\ref{eq:multi_chi-2}) and (\ref{eq:multivariate_gauss}) into the previous equation, and inverting the order of integration leads to

\begin{equation}\label{eq:Z}
Z_n(\mathbf{u})=\left(\frac{1}{(2\pi)^{n/2}\sqrt{\det(\mathbf{G})}}\int d\mathbf{X}\exp -\frac{1}{2}\mathbf{X}^T (\mathbf{G}^{-1}+2\mathbf{U})\mathbf{X}\right)^M=\frac{1}{\det(\mathbf{I}+2\mathbf{G}\mathbf{U})^{M/2}},
\end{equation}

\noindent where $\mathbf{X}=(X_1,\ldots, X_n)^T$, and $U_{ij}=u_i\delta_{ij}$. This result is similar to that obtained for the Wishart matrix distribution \cite{Kotz}. The next step is now to calculate the derivative of $Z_n$ with respect to $u_k$. We set $\mathbf{A}=\mathbf{I}+2\mathbf{G}\mathbf{U}$, and make use of the Jacobi formula

\begin{equation}
d(\det(\mathbf{A}))=\det(\mathbf{A})\textrm{Tr}(\mathbf{A}^{-1}d\mathbf{A}),
\end{equation}

\noindent where $d\mathbf{A}$ is the variation of the matrix A when $u_k$ undergoes a variation $du_k$. From there one obtains

\begin{equation}\label{eq:dZ}
\frac{\partial Z_n}{\partial u_k}(\mathbf{u})=-\frac{M}{2}\frac{\left(1-A^{-1}_{kk}\right)}{u_k}Z_n(\mathbf{u}).
\end{equation}

\noindent The second derivative can be calculated upon noting that $(A^{-1})_{kk}=\mathcal{C}_{kk}/\det(\mathbf{A})$, where $\mathcal{C}_{kk}$ is the $k^{th}$ element on the diagonal of the cofactor matrix of $\mathbf{A}$. By construction, $C_{kk}$ does not depend on $u_k$, and the following result is readily obtained 

\begin{equation}\label{eq:dZ2}
\frac{\partial^2 Z_n}{\partial u_k^2}(\mathbf{u})=\left(1+\frac{2}{M}\right)\left[\frac{M}{2}\frac{\left(1-A^{-1}_{kk}\right)}{u_k}\right]^2 Z_n(\mathbf{u}).
\end{equation}
 
\noindent As a check on these results, $\langle\nu_k\rangle$ and $\langle\nu_k^2\rangle$ can be recomputed by setting $\mathbf{u}=0$ respectively in Eq. (\ref{eq:dZ}) and Eq. (\ref{eq:dZ2}). For small $\mathbf{u}$, $\mathbf{A}^{-1}\simeq \mathbf{I}-2\mathbf{G}\mathbf{U}$ (first term of the Neumann series), so that $\langle\nu_k\rangle=MG_{kk}$
and  $\langle\nu_k^2\rangle=2MG_{kk}^2+(MG_{kk})^2$ are recovered.

\subsection{Neutral density and ionization source}

\begin{figure}[htb] 
\includegraphics[width=.45\textwidth]{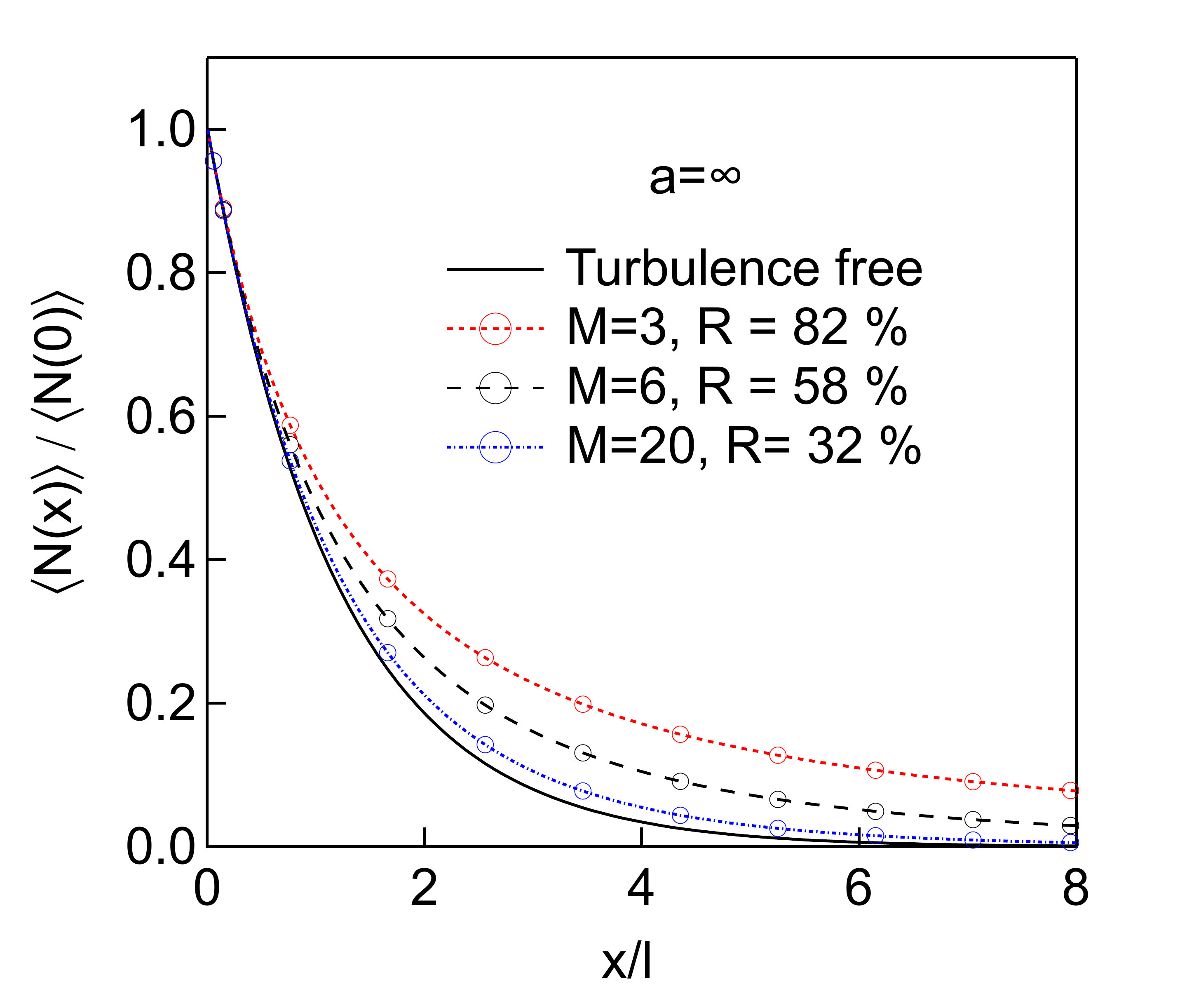}~a)
\includegraphics[width=.45\textwidth]{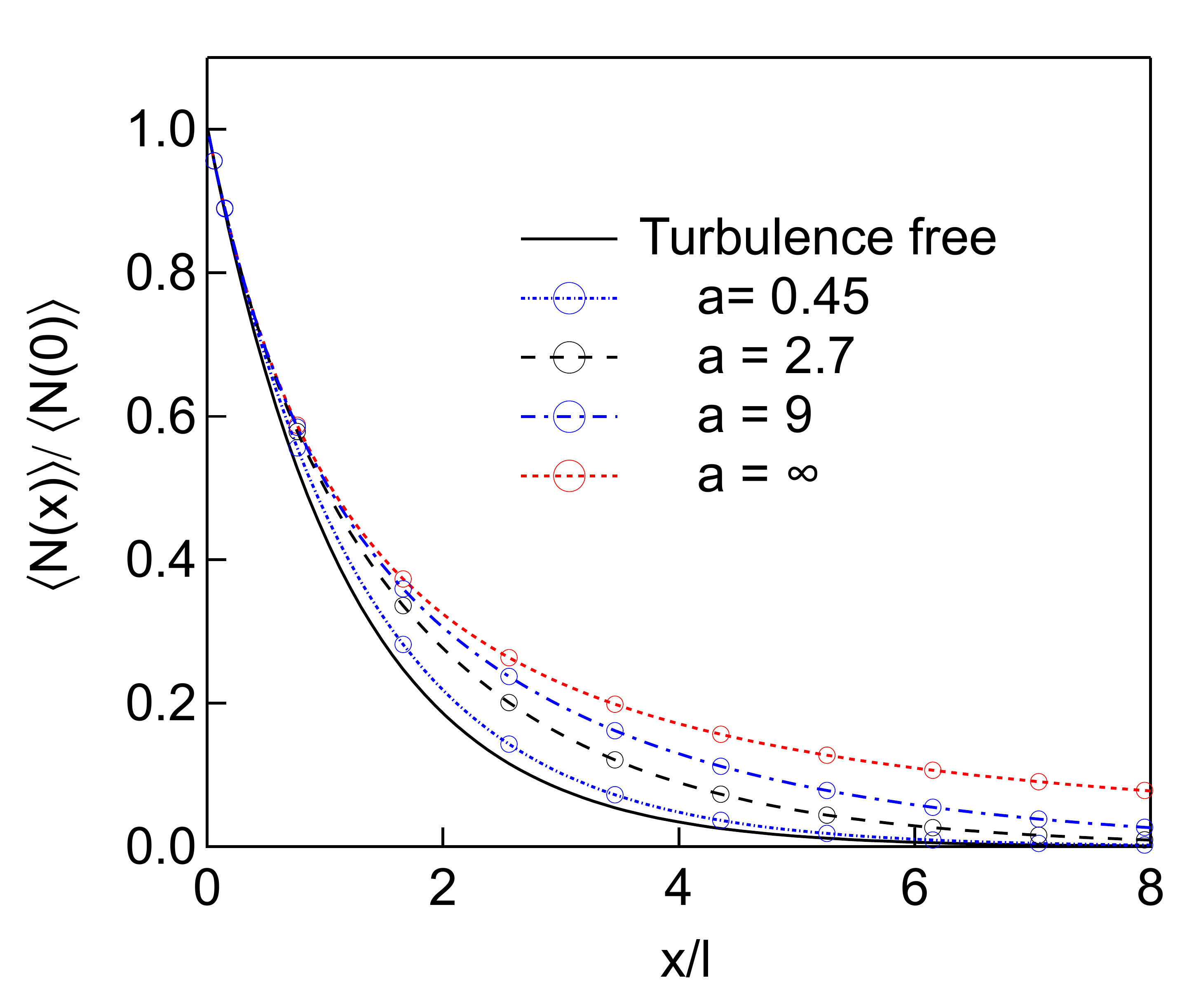}~b)
\caption{a) Average density profiles obtained with a non stochastic boundary condition, as a function of the distance of the source in units of mean free path $l$, for different values of the fluctuation rate. All the profiles are calculated for $a=+\infty$. The solid line is the turbulence free profile, the dash dotted line corresponds to R=32\%, the dashed line to R=58\%, 	and the dotted line to R=82\%. As R increases, the decay of the average density profiles becomes slower and slower.   b) Average density profile obtained for R=82 \%, for different values of the ratio of the integral scale to the mean free path a. The solid line is the turbulent free profile, the short-dotted line corresponds to a=0.45, the dashed line to a=2.7, the dash-dotted line to a=9 and the dotted line to a=$+\infty$. The larger a, the slower the neutral particle density decays. On both plots, open circles represent the results of the Monte Carlo simulations carried out in section \ref{sec:num}. \label{fig:av_dens}}
\end{figure}

\begin{figure}[htb]
\includegraphics[width=.45\textwidth]{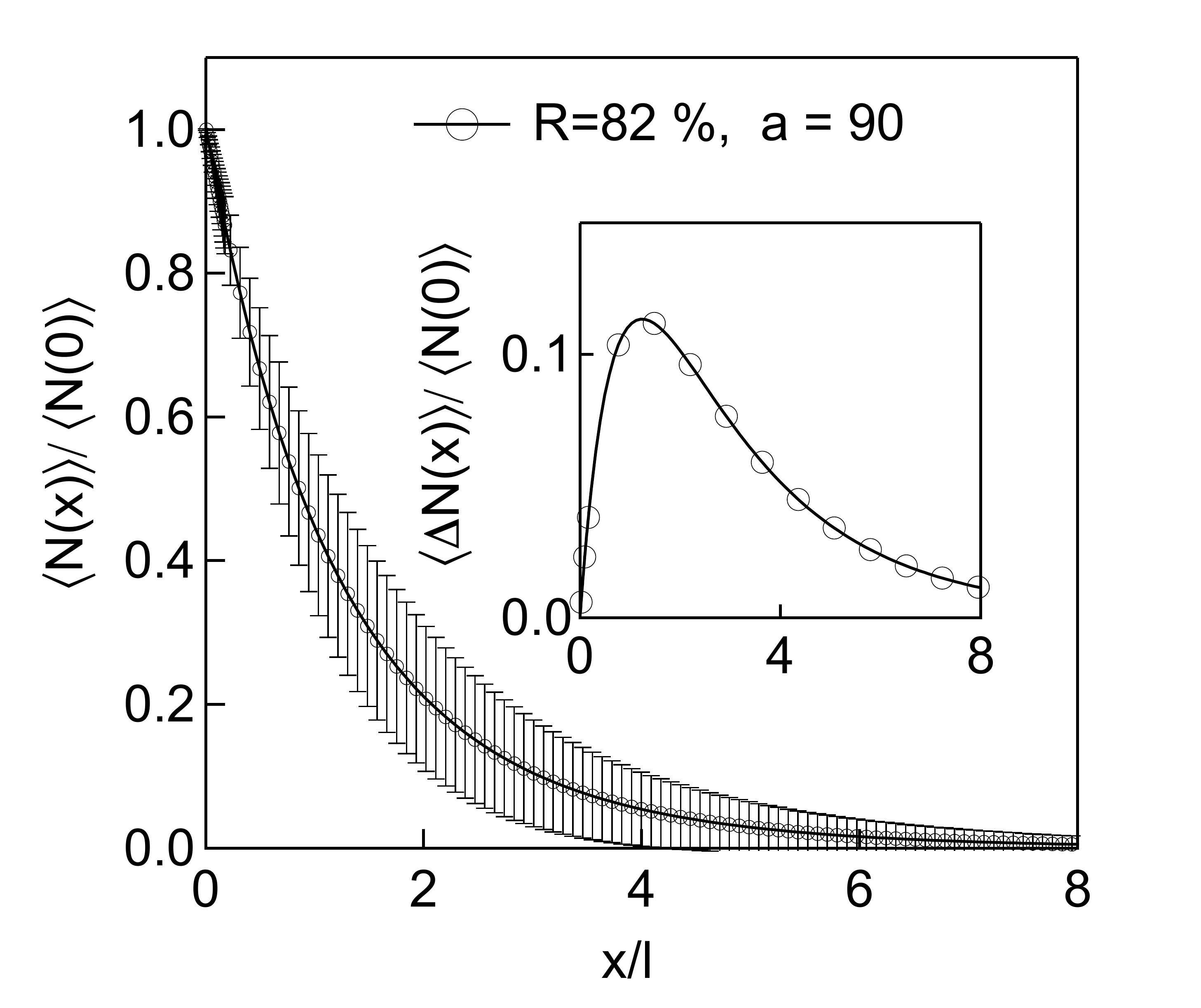}~a)
\includegraphics[width=.45\textwidth]{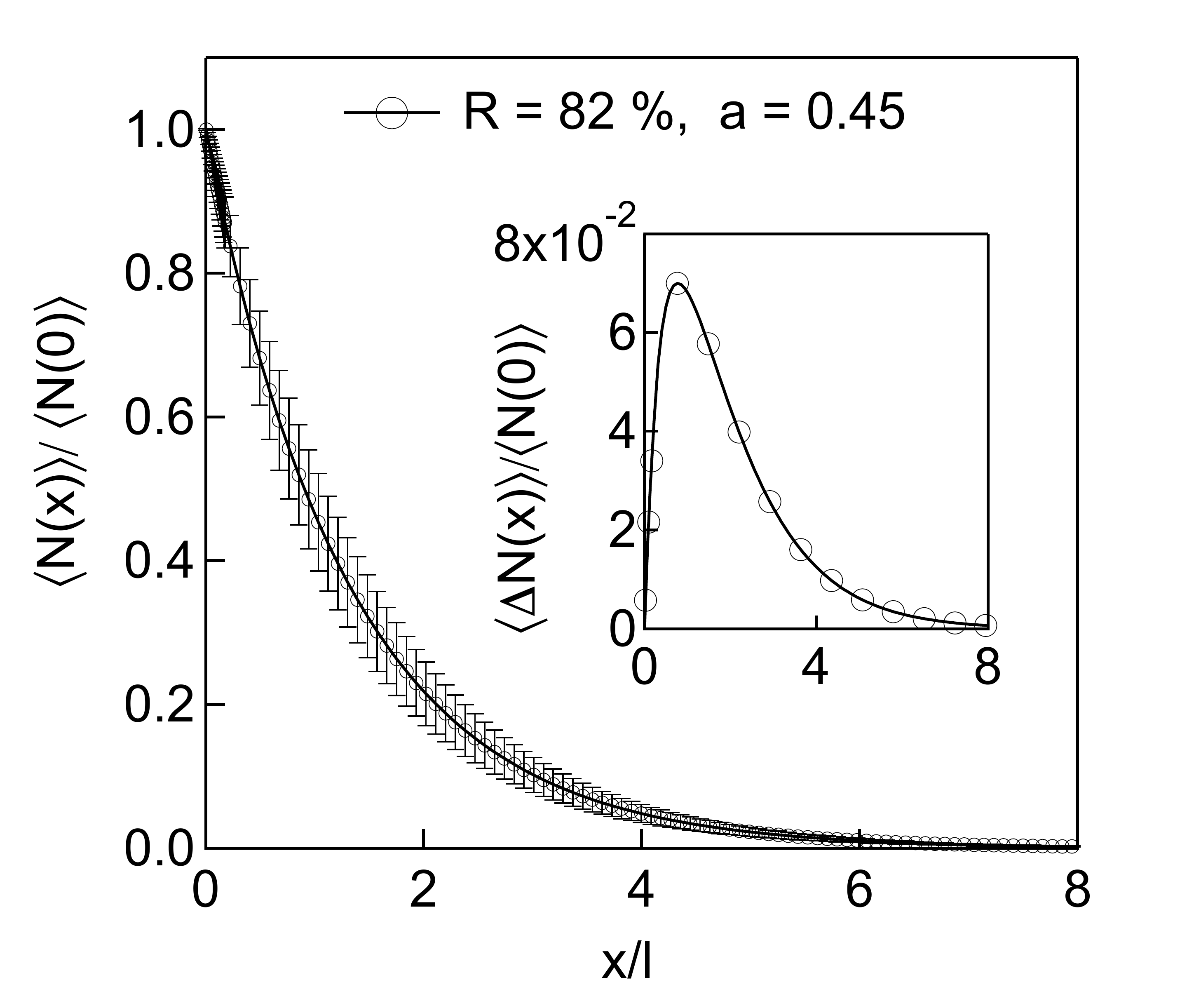}~b)
\caption{a) Plot of the average density profile as a function of the distance to the wall in units of mean free path l, for R=82\% and $a=90$. The error bars are such that $\langle N\rangle \pm\langle\Delta N\rangle$, where $\langle\Delta N\rangle$ is the standard deviation profile, which represents the typical statistical dispersion of the density profiles. $\langle\Delta N\rangle$ is plotted in the inset, normalized to the value of the density at the wall. For $x=0$, $\langle\Delta N\rangle=0$ because of the non stochastic boundary condition.  b)  Same as a), but for a=0.45. The dispersion is clearly lower by a factor of about 2 compared to a). This result is consistent with what is expected from the ergodic theorem, which implies $\langle\Delta N\rangle\rightarrow 0$ as $a\rightarrow 0$. On both plots, open circles represent the results of the Monte Carlo simulations carried out in section \ref{sec:num}.\label{fig:std_lambda}}
\end{figure}

\begin{figure}[htb]
\includegraphics[width=.45\textwidth]{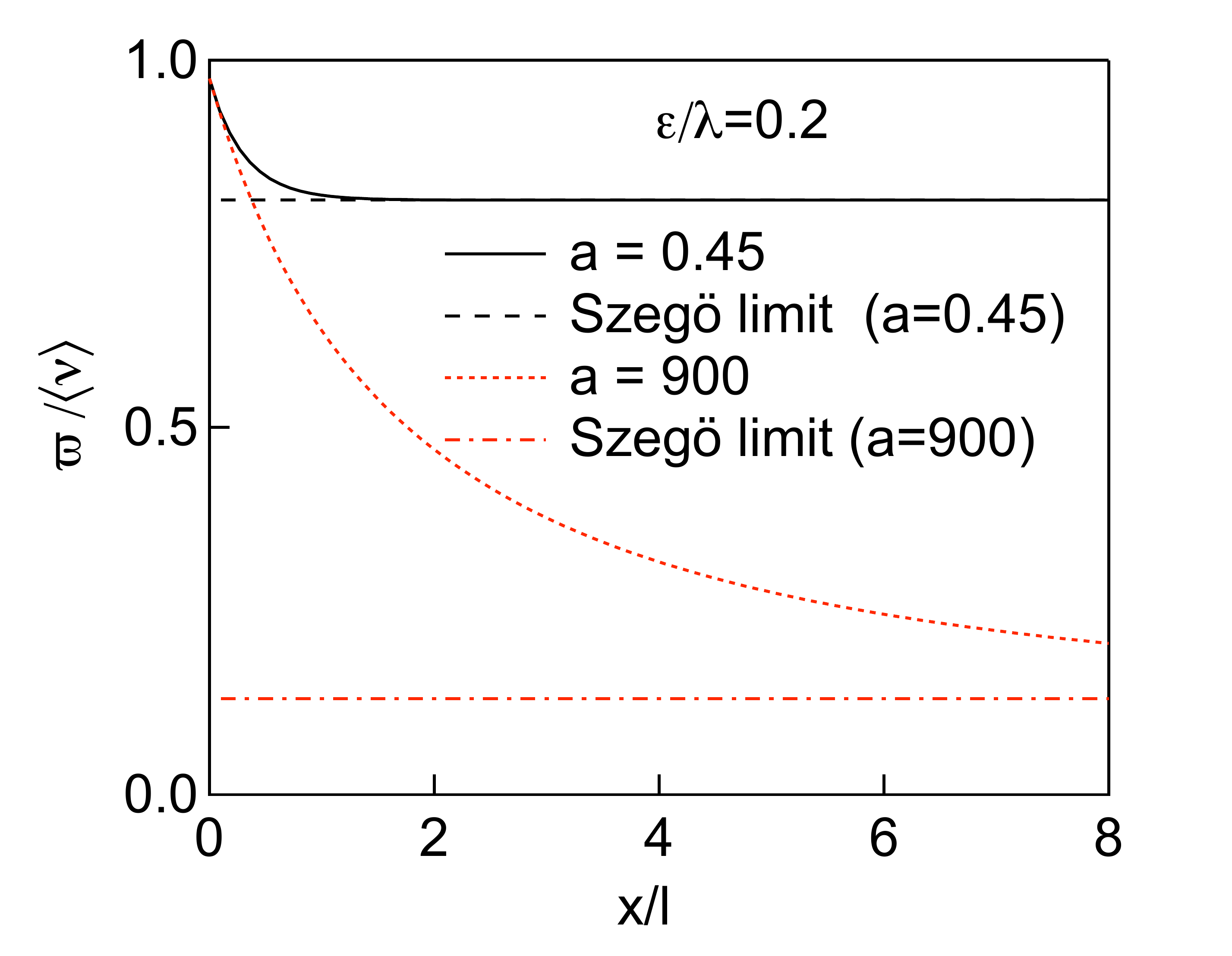}
\caption{Plot of the effective ionization rate $\varpi$ in units of $\langle\nu\rangle$, as a function of the distance to the source for two values of the integral scale, namely a=0.45 and a=900 (in units of l). For a=0.45, $\varpi(x)$ decays towards a constant value of the order of 0.8, in accordance with Szeg\"{o}'s theorem prediction, Eq. (\ref{eq:discrete_Szego}). In practice, this value is reached after a distance of the order of twice the integral scale. \label{fig:effective_nu}}
\end{figure}

\begin{figure}[htb]
\includegraphics[width=.45\textwidth]{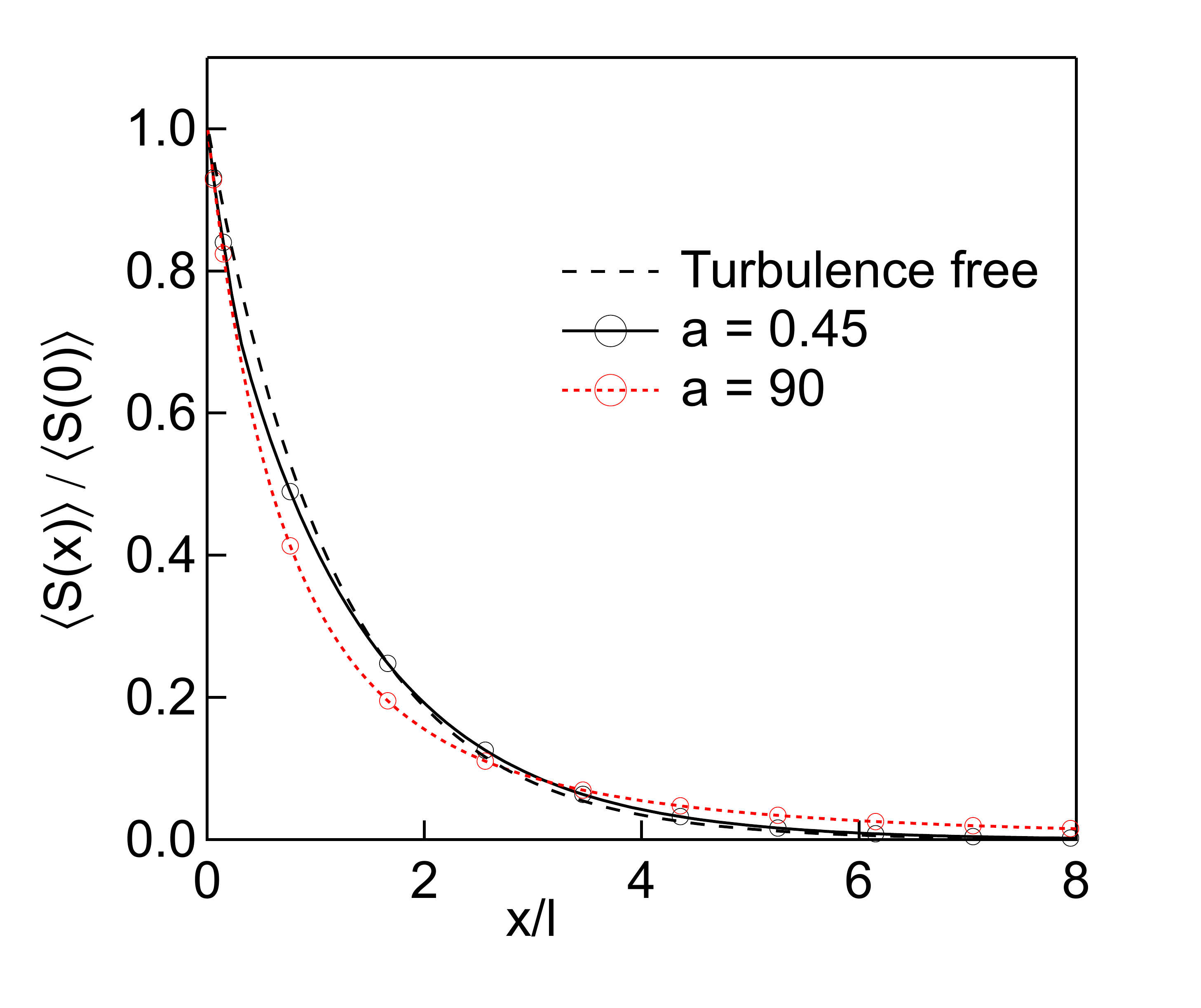}~a)
\includegraphics[width=.45\textwidth]{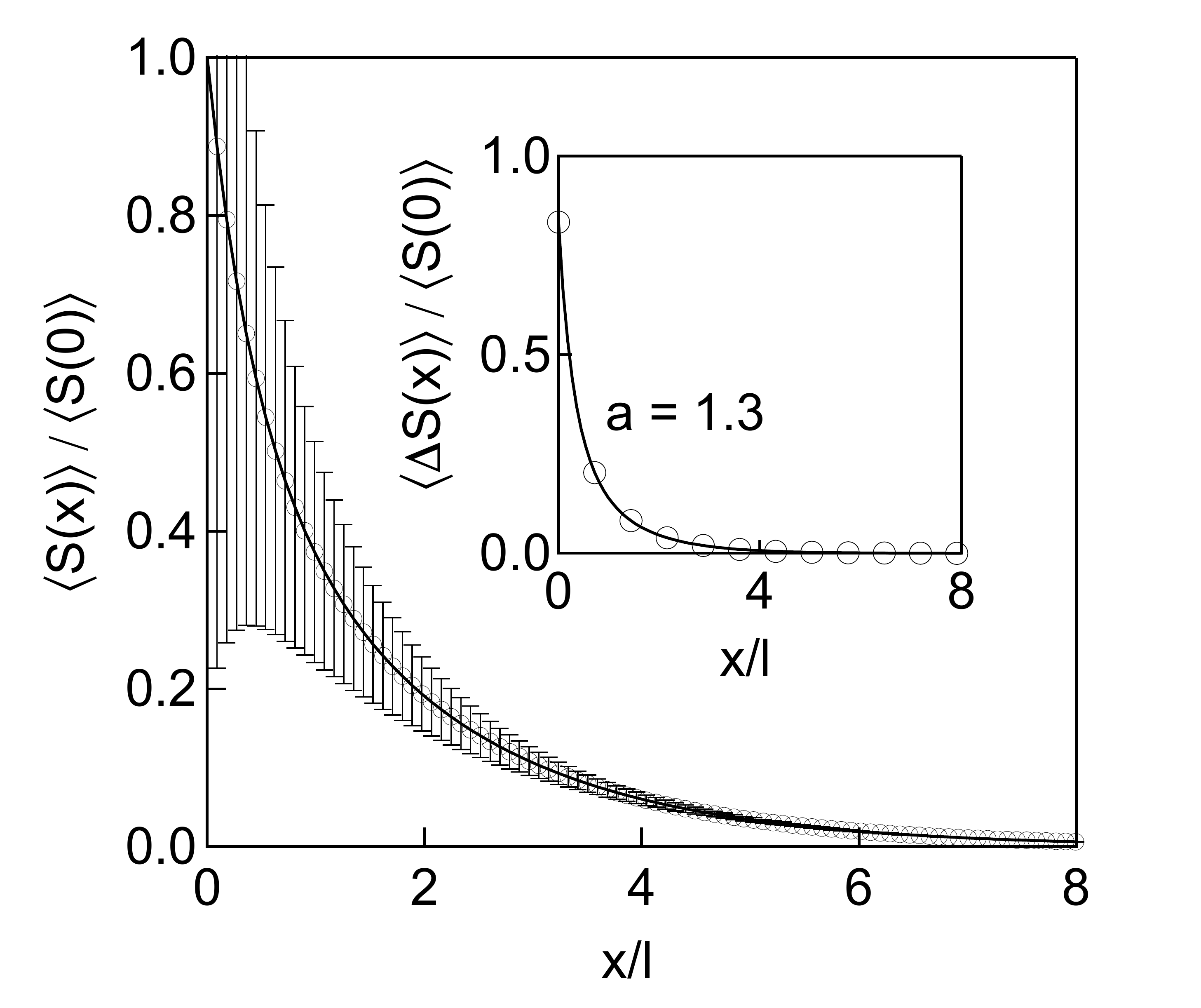}~b)
\caption{a) Plot of the average ionization source $\langle S\rangle$ as a function of the distance to the wall in units of the mean free path, for $R=82$\%. The dashed line is the turbulence free case, the solid line corresponds to $a=0.45$, while for the dotted line $a=90$. The differences between the averaged and the turbulent free profiles are smaller than for the average density. b) Plot of the average ionization source, for R=82\% and a=1.3, with errors bars representing the standard deviation $\langle\Delta S\rangle$. In contrast with the neutral particle density, the dispersion is the largest at the wall. Open circles represent the results of the Monte Carlo simulations carried out in section \ref{sec:num}. \label{fig:Sio}}
\end{figure}

The results obtained in the previous section, namely Eqs. (\ref{eq:Z}) and (\ref{eq:dZ}), provide analytical expressions for the average neutral density and ionization source when the flux at the wall $\Gamma_0$ is non-stochastic (slow recycling). The $p^{th}$ moment of the density is given by

\begin{equation}\label{eq:pth_moment}
\langle N^p(x_k)\rangle = \left(\frac{\Gamma_0}{v_0}\right)^p Z_n(p\mathbf{\mathbf{u}_k})=\left(\frac{\Gamma_0}{v_0}\right)^p\frac{1}{\det(\mathbf{I}+2p\mathbf{G}\mathbf{U}_k)^{M/2}},
\end{equation}

\noindent where $(U_k)_{ij}=u_k\delta_{ij}=\epsilon/v_0\delta_{i\leq k} \delta_{ij} $. This result can be validated by investigating several limiting cases (see Appendix B). The average density ($p=1$) is plotted on Fig. \ref{fig:av_dens} as a function of the distance to the source in units of the mean free path $l$, for different values of the ratio $a=\lambda/l$. As an example, for $N_e=5\times10^{18}$ m$^{-3}$, $T_e=50$ eV and $E_0=mv_0^2/2=20$ eV ($m$ is the mass of the atom), the ionization mean free path $l$ is of the order of 20 cm for a D atom. In the same conditions, for a D$_2$ molecule with $E_0=0.03$ eV, $l \simeq 8 $ mm.
Fig. \ref{fig:av_dens}-a) shows the average density in the $\lambda\rightarrow\infty$ case, for different values of the fluctuation rate $R$. It is seen that the average density profiles decrease more slowly than the fluctuation free profile ($R=0$, solid line), in accordance with Eq. (\ref{eq:Jensen}), but that significant deviations occur only for large fluctuation rates above 50 \%. Fig. \ref{fig:av_dens}-b) shows that as $\lambda$ tends to zero, the average density profile comes closer and closer to the turbulence free profile, as expected from the arguments developed in section A. The slowest decaying profile corresponds to $a\rightarrow \infty$. The profiles for finite values of $a$ are close to this limiting case provided $x \ll \lambda$. On the size $L=8 \ l$ of the box, $a=90$ provides an excellent approximation to the $a=\infty$ case. The standard deviation of the density is obtained from Eq. (\ref{eq:pth_moment}) by $\langle\Delta N(x_k)\rangle=\left(\langle N^2(x_k)\rangle-\langle N(x_k)\rangle^2\right)^{1/2}$, and is plotted on Fig. \ref{fig:std_lambda}-a) for $a=90$ and b) for $a=0.45$ and $M=3$, both as error bars on the average density profile and in the inset. The latter represents the typical dispersion of the different realizations of the density profile, and decreases with $\lambda$ in accordance with the  results of section A (ergodic theorem).
The standard deviation is zero for $x=0$, because the boundary condition prescribed here is non stochastic. The average ionization source and its second moment are respectively given by

\begin{equation}\label{eq:Sio_av}
\langle S(x_k)\rangle = -\frac{\Gamma_0}{v_0}\frac{\partial Z}{\partial u_k}(\mathbf{\mathbf{U}_k})= \varpi(x_k) \langle N(x_k)\rangle ,
\end{equation}

\begin{equation}\label{eq:Sio_std}
\langle S^2(x_k)\rangle = \left(\frac{\Gamma_0}{v_0}\right)^2\frac{\partial^2 Z}{\partial u_k^2}(\mathbf{\mathbf{U}_k})=\left(1+\frac{2}{M}\right)\varpi^2(x_k)\langle N^2(x_k)\rangle ,
\end{equation}

\noindent with

\begin{equation}
\varpi(x_k) = \frac{Mv_0}{2} \frac{(1-(A_k^{-1})_{kk})}{\epsilon},
\end{equation}
  
\noindent where we recall that $\mathbf{A}_k=\mathbf{I}+2\mathbf{G}\mathbf{U}_k$. $\varpi(x_k)$ is an effective ionization rate, which depends on space. Its asymptotic behavior for large values of $x_k$ is obtained from Szeg\"o's theorem on Toeplitz matrices \cite{Gray} in Appendix \ref{app:Szego}. In the continuum limit ($\epsilon\rightarrow 0$), and for the correlation function given by Eq. (\ref{eq:correl_function_exp}), we have

\begin{equation}\label{Eq:Szego_continuous}
\varpi_{\infty}=\lim_{k\rightarrow+\infty}  \varpi(x_k) = \frac{\left(1+2\langle\nu\rangle\tau\right)^{1/2}-1}{\tau},
\end{equation}

\noindent  where the time $\tau$ is such that $\tau=4\lambda/Mv_0$. Eq. (\ref{Eq:Szego_continuous}) shows that $\langle S\rangle$ and $\langle N\rangle$ have the same asymptotic decay, except in the infinite correlation length case where $\varpi_{\infty}$ tends to 0. The effective ionization rate $\varpi$ is plotted on Fig. \ref{fig:effective_nu}, for $a=0.45$ and $a=900$, together with the corresponding asymptotic values from Szeg\"o's theorem. This value is reached at a distance of a few correlation lengths from the wall, where $\varpi(x_1)=\langle\nu\rangle$. In the $a=0.45$ case, $\epsilon/\lambda=0.2$, and the difference between the continuum limit and the discrete result given in appendix C (Eq. (\ref{eq:discrete_Szego})) is of the order of 2\%. The average ionization source $\langle S(x_k)\rangle$ is plotted on Fig. \ref{fig:Sio} for $a=0.45$ and $a=900$.
The differences between the turbulence free and the averaged ionization source are much smaller than in the case of the neutral particle density, because the enhanced neutral particle penetration originating from low density realizations is compensated by the corresponding low ionization rates.  The standard deviation is this time maximum at the wall, since the fluctuations of $S(0)$ directly reflects those of $N_e(0)$. This can be checked directly by forming $\langle S^2(0)\rangle-\langle S(0)\rangle^2$ from Eq.  (\ref{eq:Sio_av}) and (\ref{eq:Sio_std}). Finally, we come back to the neutral particle density, this time when the boundary condition $\Gamma_0$ is stochastic (fast recycling, denoted by the $^F$ superscript). In our model, the recycling flux is directly proportional to the plasma density, \textit{i.e.} to the ionization rate, and the average neutral particle density and its second moment are given by

\begin{equation}\label{eq:Nav_rec}
\langle N^{F}(x_k)\rangle = -\frac{\langle\Gamma_0\rangle}{v_0\langle \nu\rangle}\frac{\partial Z}{\partial u_1}(\mathbf{{U}_k})=\frac{\varpi(x_k)}{\langle\nu\rangle}\langle N(x_k)\rangle
\end{equation}

\begin{equation}\label{eq:Nsq_rec}
\langle N^{F}(x_k)^2\rangle =\left(1+\frac{M}{2}\right)\left(\frac{\varpi(x_k)}{\langle\nu\rangle}\right)^2\langle N^2(x_k)\rangle
\end{equation}

\noindent according to Eq. (\ref{eq:dZfunc}), (\ref{eq:dZ}), and (\ref{eq:dZ2}) and using $(A_k^{-1})_{11}=(A_k^{-1})_{kk}$ (see Appendix \ref{app:Szego}). Therefore, the results are the same as for the ionization source, up to a dimensional factor. In this case, the source strength plays the same role as the ionization rate for S in the slow recycling case.\\

The results presented in this section highlight the essential role played by the ratio $a$ of the turbulence correlation length to the neutral mean free path. This provides an important clue to understand 2D and scattering effects.  Moreover, if the monokinetic approximation is relaxed, linearity implies that the average density can be obtained by further averaging Eq. (\ref{eq:pth_moment}) over the initial velocity $v_0$ distribution function. This is equivalent to average over $a$, since the neutral mean free path is directly proportional to  $v_0$, so that in the end one might have properly scaled contributions from both $a\ll 1$ and $a\gg 1$ regimes. For instance, atoms created by molecular dissociation in edge plasmas can have energies as low as 0.2 eV, while backscattering at the wall leads to atoms having energies of several hundreds of eV.

\subsection{Influence of the correlation function shape}

\begin{figure}[htb]
\includegraphics[width=.45\textwidth]{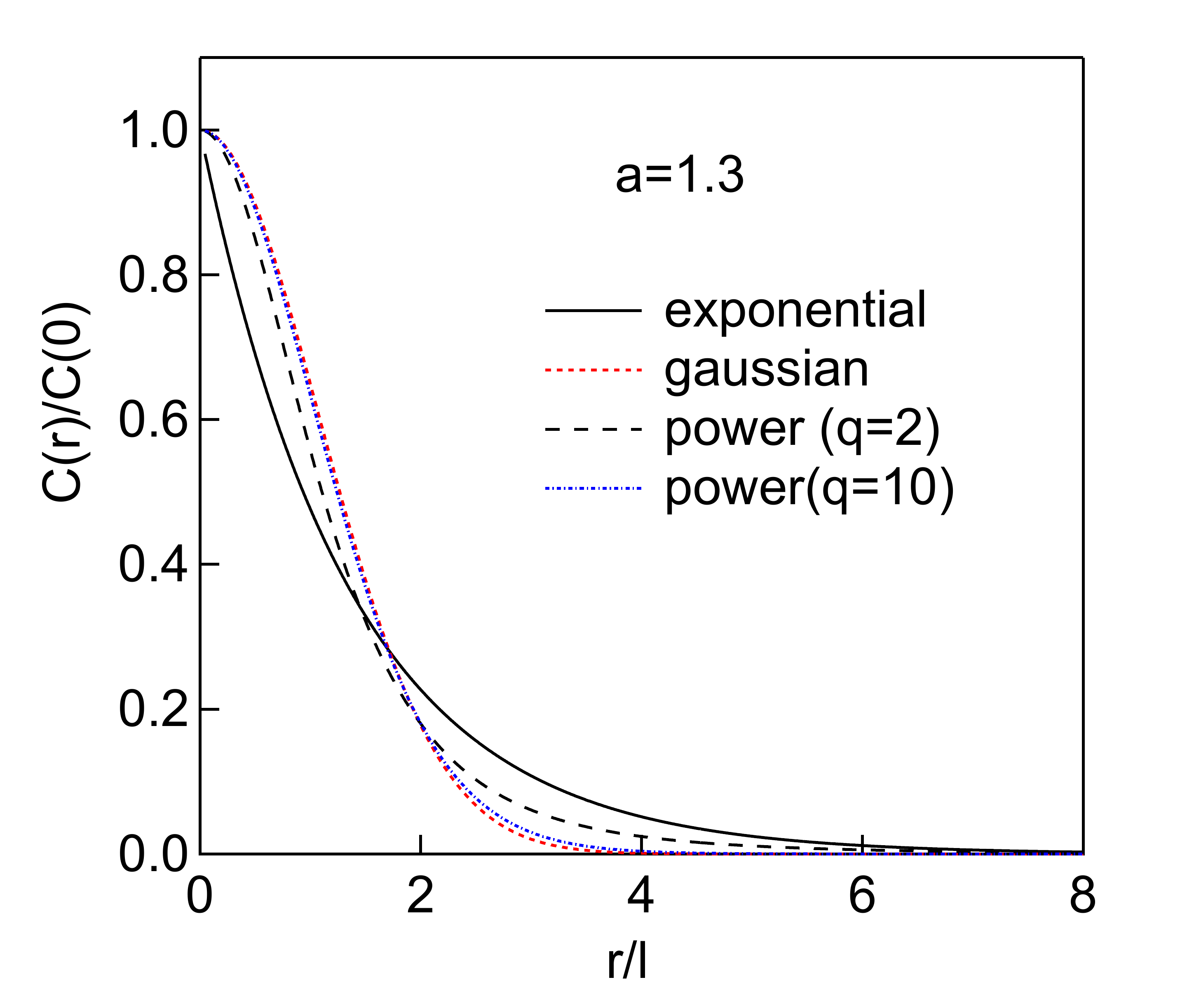}~a)
\includegraphics[width=.45\textwidth]{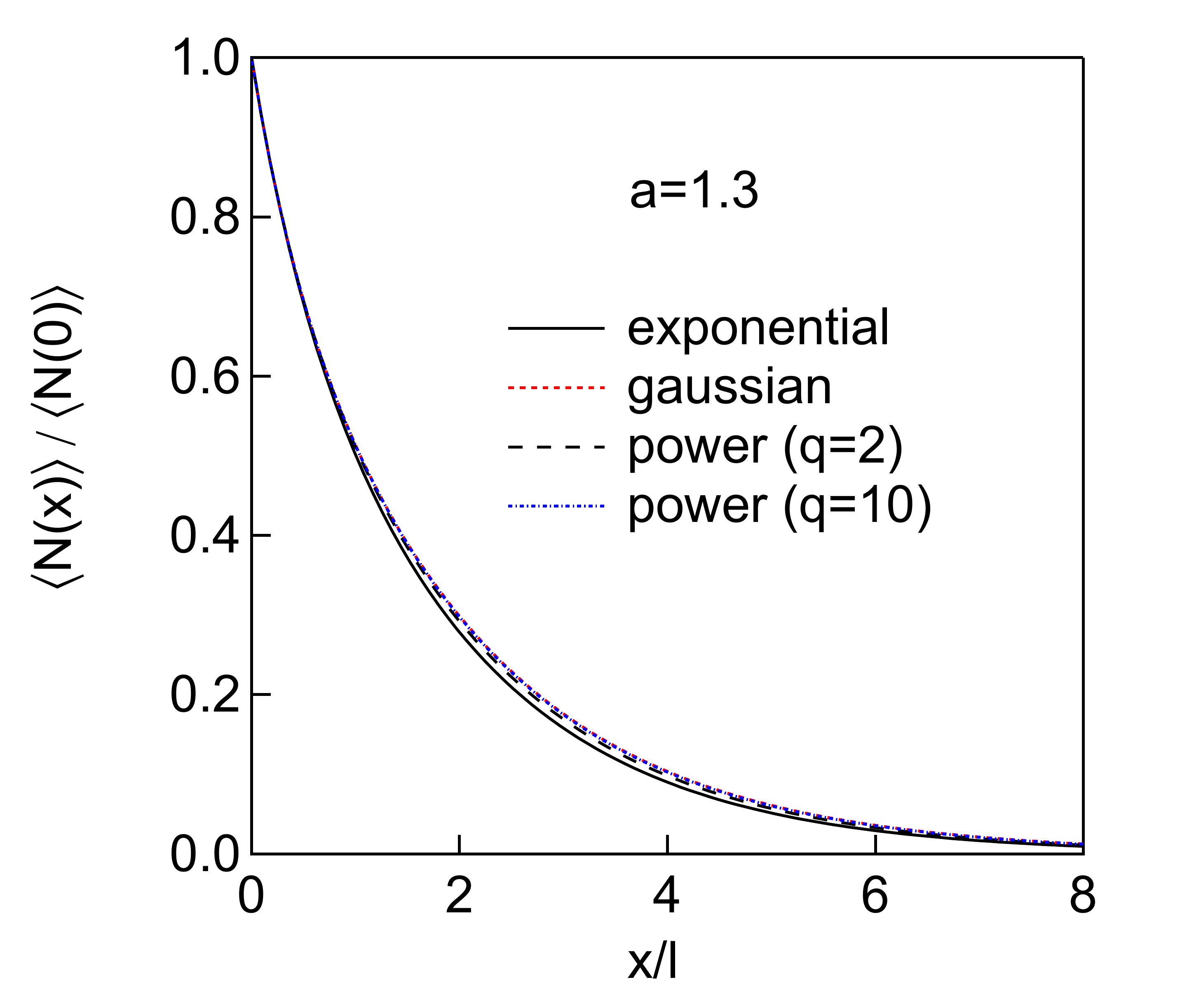}~b)
\caption{a) Plot of an exponential (Eq. (\ref{eq:correl_function_exp}), solid line), gaussian (dotted line), and power law (Eq. (\ref{eq:Cnr}) for q=2, dashed line, and for q=10, dash-dotted line) correlation functions, with the same integral scale (\textit{i.e.} the same area under the curve) such that a=1.3. b) Plot of the corresponding average density profiles as a function of the distance to the source in units of the mean free path. The different profiles are found to be very similar. \label{fig:Cr_shape}}
\end{figure}

In the previous section, the role of the integral scale $\lambda$ has been discussed for a correlation function $C(r)$ of the form given by Eq. (\ref{eq:correl_function_exp}). However, Eq. (\ref{eq:pth_moment}) is valid for any positive (a restriction of the multivariate Gamma model) $C(r)$. In this section, we address the sensitivity of the average density profile to the actual expression of $C(r)$, for the same value of the integral scale $\lambda=\int_0^{+\infty} C(r)dr/C(0)$. We first consider the following family of correlation functions $C_q(r)$

\begin{equation}\label{eq:Cnr}
C_q(r)=\frac{\sigma^2}{(1+(r/r_0)^{2})^q},
\end{equation}

\noindent for which $r_0=(2\lambda/\pi) (2q-2)!!/(2q-3)!!$. In addition, we consider a Gaussian correlation function $C_g(r)$, which decays faster than Eq. (\ref{eq:correl_function_exp}). These various functions are compared on Fig. \ref{fig:Cr_shape} a), with $q=1$ and $q=10$ for $C_q$, for the same value of $a=1.3$. The resulting density profiles are plotted on Fig. \ref{fig:Cr_shape}  b), and are found very close from each other. This observation is found to be valid for any value of a, the largest deviations being found for $a\rightarrow +\infty$, where the effect of fluctuations is the largest (for $a\rightarrow 0$ all the density profiles tend to the fluctuation free result). Therefore, for a given integral scale, the shape of the correlation function is likely to have only a minor effect on the average quantities. In other words, the main control parameter is the integral scale of the fluctuations.

\subsection{2D effects}

\begin{figure}[htb]
\includegraphics[width=.45\textwidth]{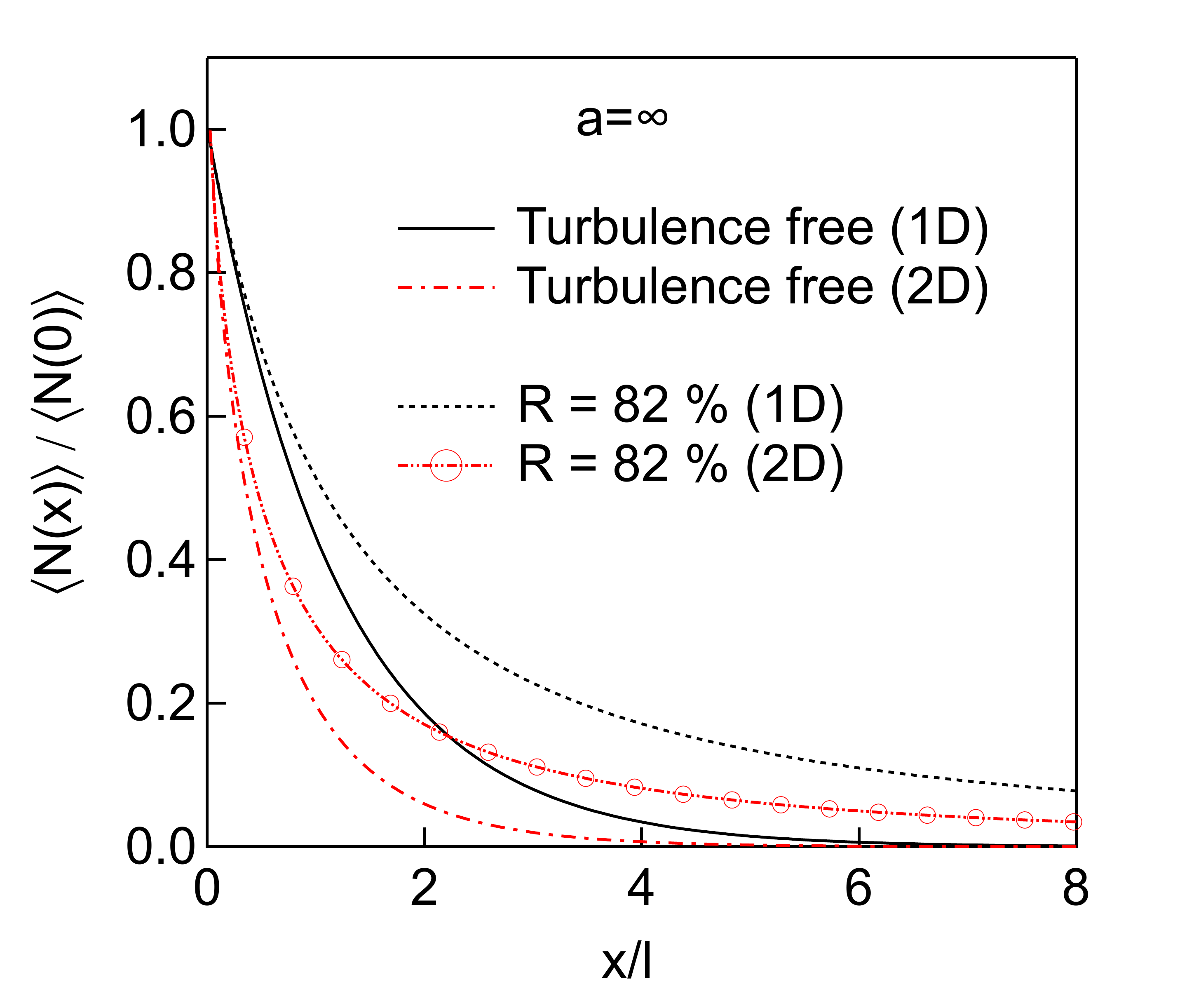}
\caption{Comparison between the 2D and the 1D cases for $M=3$ and $a=\infty$. The solid line and the dash-dotted lines are the turbulence free profiles in 1D and 2D respectively. The dotted and the dashed-double dotted line are the average density profiles in 1D and 2D respectively. The effect of fluctuations is similar in both cases. Open circles represent the results of the Monte Carlo simulations carried out in section \ref{sec:num}. \label{fig:2Dvs1D}}
\end{figure}

We now investigate 2D effects, since Monte Carlo calculations including scattering (\textit{i.e.} charge exchange) will be performed in 2D. The neutral density profile is expected to be more peaked towards the wall in 2D than in 1D, because neutrals which have large $v_y$ velocity components will be ionized closer from the wall (along $x$) than those having small or zero $v_y$ component. In fact, if $\theta$ is the angle between the initial velocity (\textit{i.e.} $\boldsymbol{\Omega}$) and (Ox), $x=s/\cos\theta$. Assuming a cosine distribution for the particle source at the wall, the following solution for $f$ in an homogeneous plasma is readily obtained by averaging Eq. (\ref{eq:Boltzmann_int}) in the scattering free limit over angles

\begin{equation}
f(x,v)=2N_0 \delta(v-v_0)\frac{\nu x}{v_0}\textrm{E}_2\left(\frac{\nu x}{v_0}\right),
\end{equation}
 
\noindent where $\textrm{E}_2$ is an exponential integral \cite{Gradshteyn}. The following result for the average density is easily obtained from Eq. (\ref{eq:pth_moment}) in the slow recycling case

\begin{equation}\label{eq:pth_moment_2D}
\langle N_{2D}(x_k)\rangle \simeq 2\left(\frac{\Gamma_0}{v_0}\right) \int_0^1 d\mu  \frac{1}{\det(\mathbf{I}+2\mathbf{G}\mathbf{U}_k/\mu)^{M/2}},
\end{equation}

\noindent where $\mu=\cos\theta$. The latter expression of $\langle N_{2D}(x_k)\rangle$ would be strictly exact in the continuum limit, since in principle $\mathbf{G}$ and $\mathbf{U}_k$ depend on $\mu$ even if isotropy (namely $C(\mathbf{r}-\mathbf{r}')=C(|\mathbf{r}-\mathbf{r}'|)$) is assumed, because of finite grid step effects. However, Eq. (\ref{eq:pth_moment_2D}) provides very good results in practice for $\epsilon/\mathcal{L}=10^{-2}$. The density profiles calculated for $M=3$ and $a=\infty$ are compared to their 1D counterparts on Fig. \ref{fig:2Dvs1D}. 
\noindent Both 2D density profiles are found to be more peaked towards the wall than in 1D. It should be noted that the angle average can be interpreted as an initial velocity average, with a velocity distribution $f_0(v)$ such that $f_0=0$ for $v>v_0$. 

\section{\label{sec:3}Implementation in the EIRENE Monte Carlo code}\label{sec:num}

We now discuss the numerical implementation of our stochastic model in the parallelized (MPI) EIRENE Monte Carlo solver \cite{Reiter05}. We have adopted the following procedure.  The plasma parameters, say the electron density, are generated by sampling $n^2$ values from the multivariate Gamma distribution
defined in section \ref{subsec:stochastic_model}, then Boltzmann's equation is solved in this plasma
background. These steps are repeated until statistical noise due to random sampling of the background medium is reduced to a negligible level, compared to the intrinsic Monte Carlo noise for each realization of the stochastic plasma fields. The numerical implementation is validated against the analytical results obtained in the previous section. The effects of scattering (\textit{i.e.} charge exchange in our context) are then finally investigated.

\subsection{Sampling from the multivariate Gamma distribution}

\begin{figure}[htb]
\includegraphics[width=.45\textwidth]{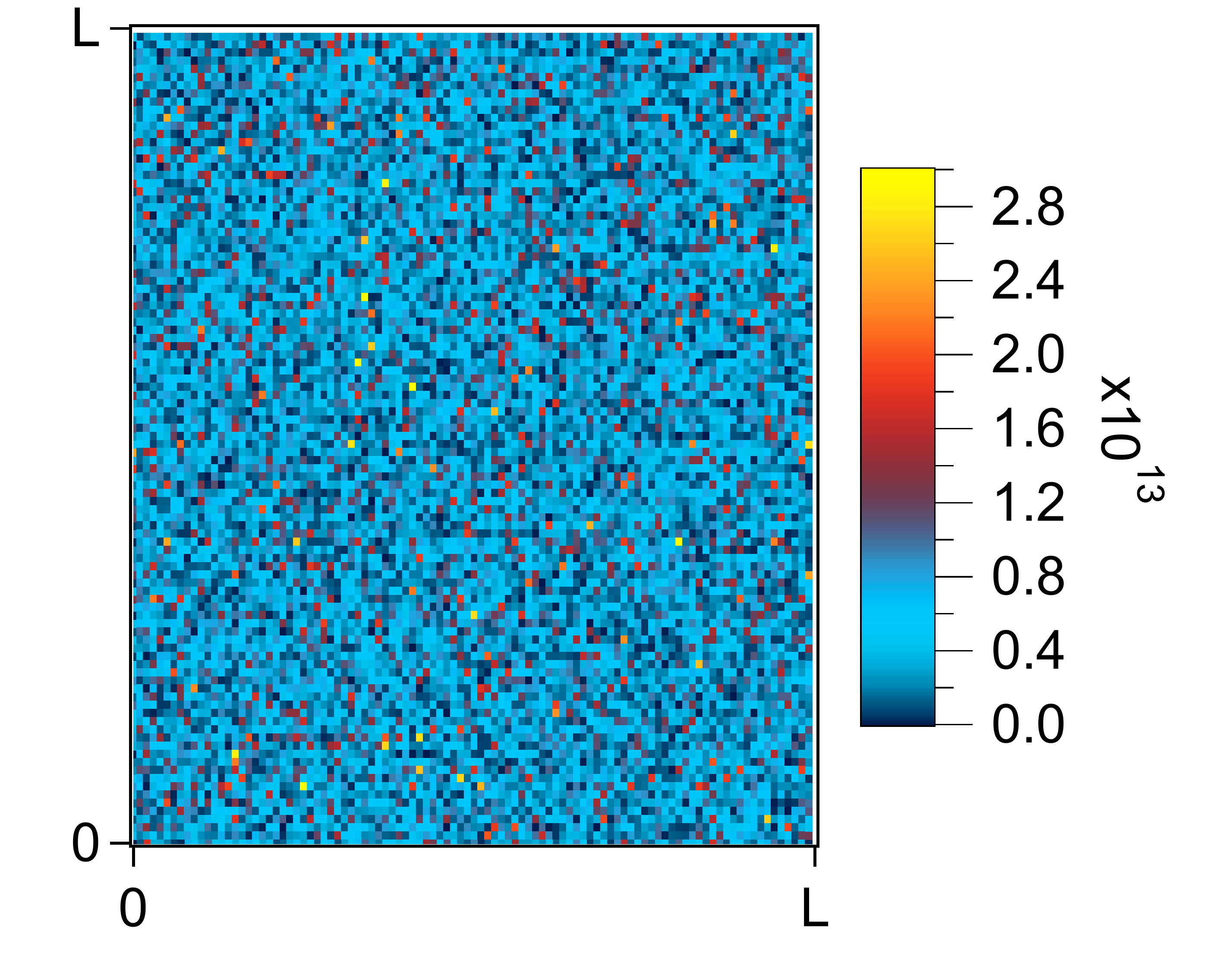}~a)
\includegraphics[width=.45\textwidth]{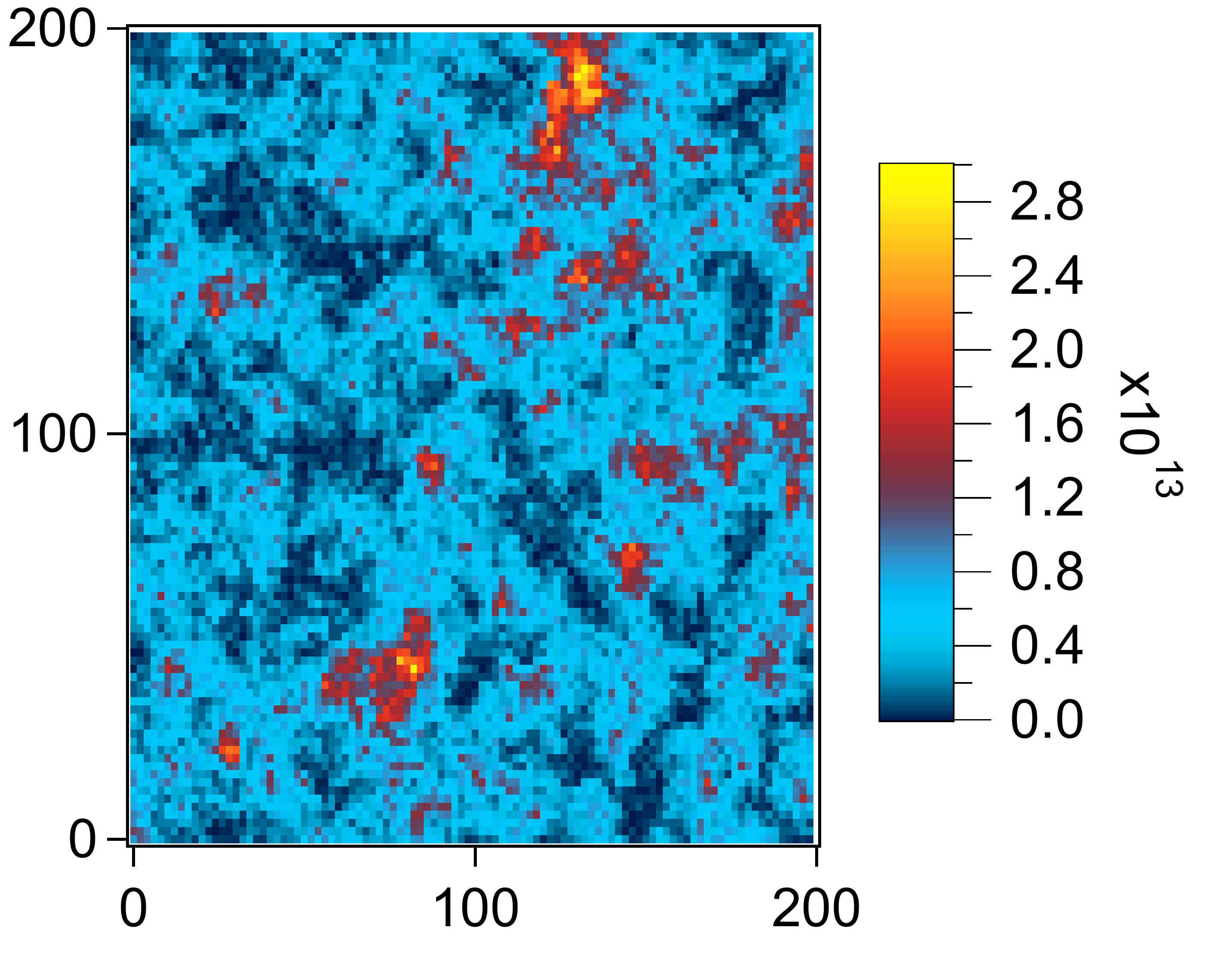}~b)
\includegraphics[width=.45\textwidth]{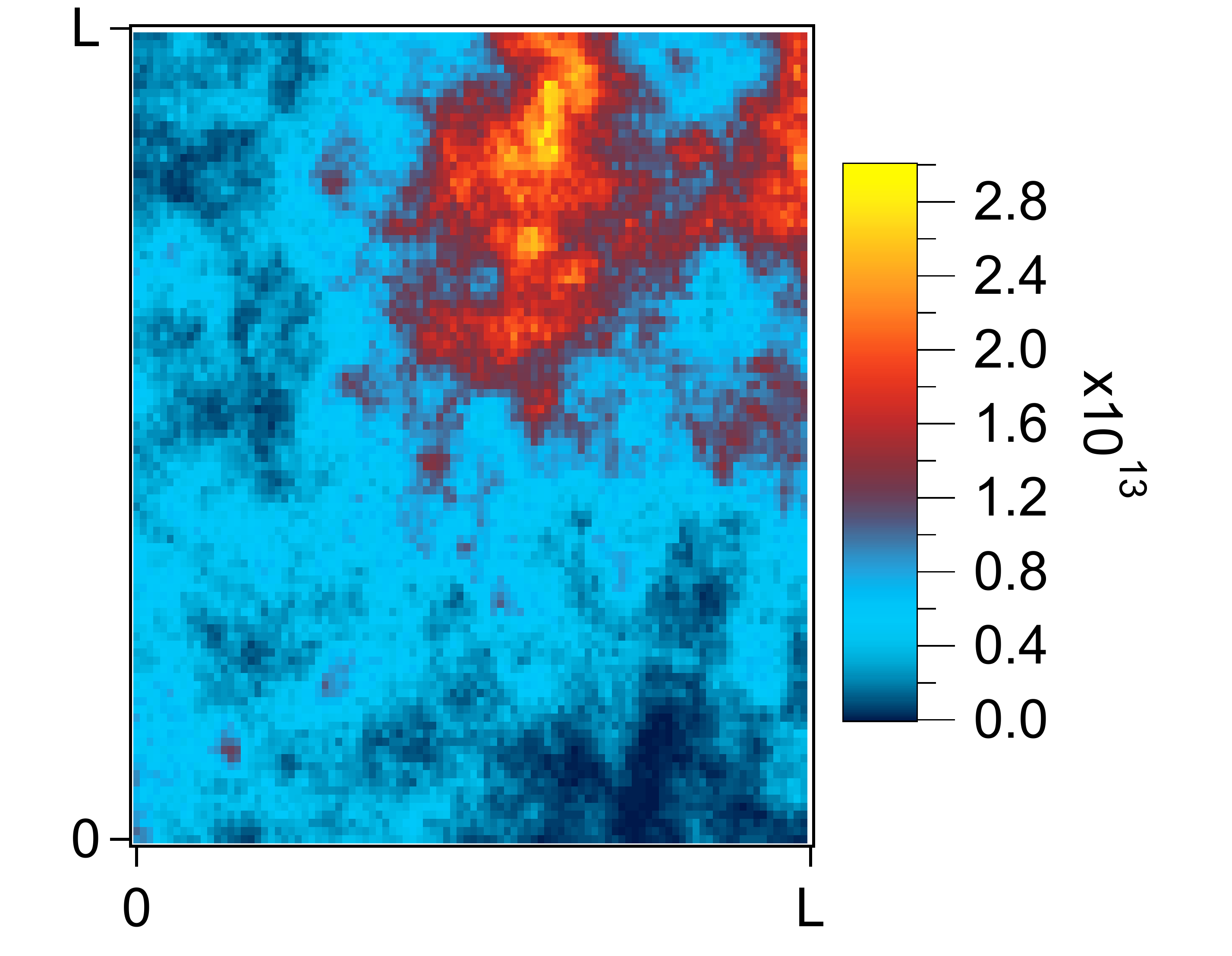}~c)
\includegraphics[width=.45\textwidth]{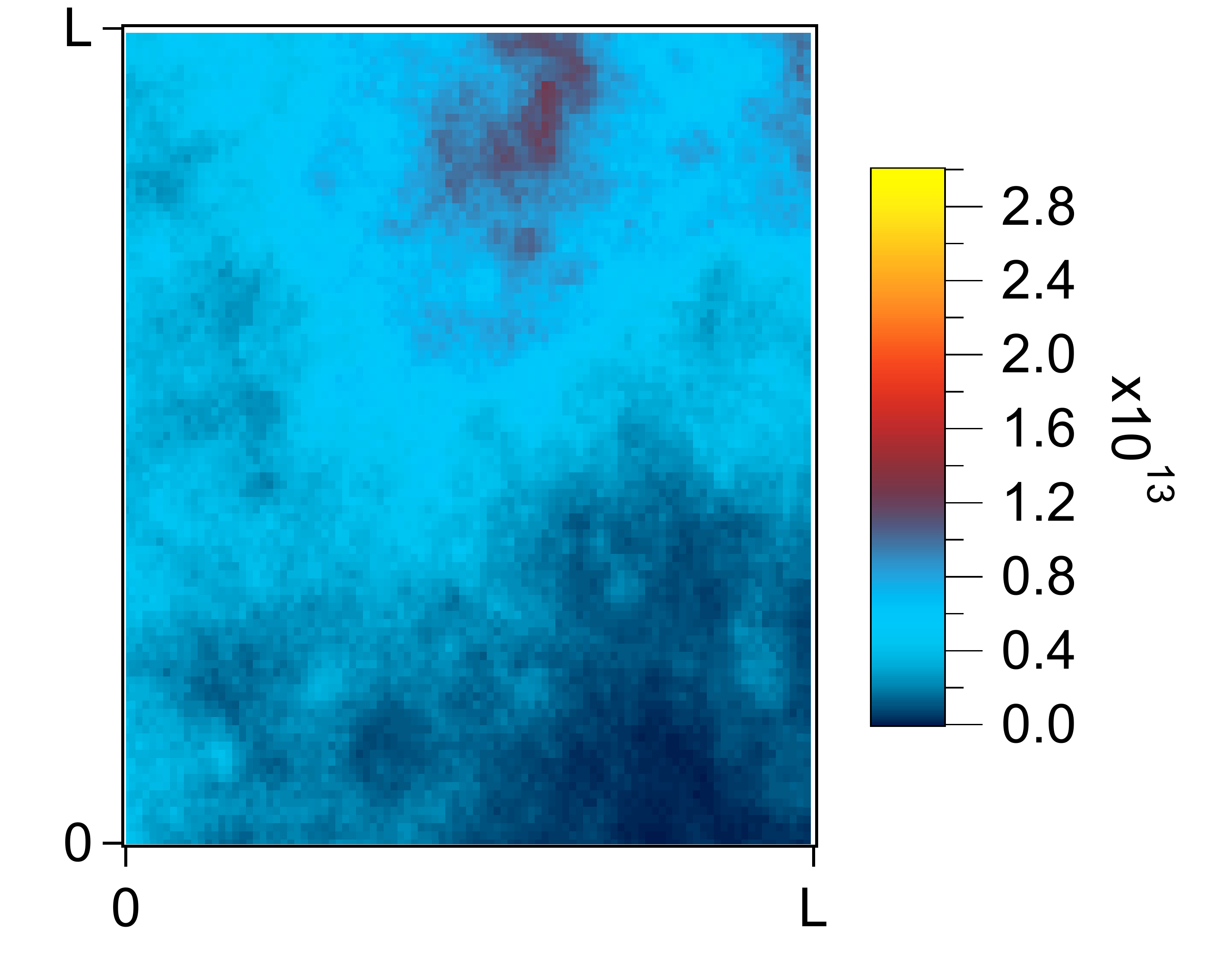}~d)
\caption{$100\times 100$ 2D map sampled from the multivariate Gamma distribution, for $M=3$ and a) $\lambda/\mathcal{L}=10^{-3} $, b) $\lambda/\mathcal{L}=2.5\times 10^{-2}$,  c) $\lambda/\mathcal{L}=0.15$ d) $\lambda/\mathcal{L}=$ 2. Note that the average density is $\langle N_e\rangle=5\times 10^{12}$ cm$^{-3}$. \label{fig:maps_ex}}
\end{figure}

From the definition of the multivariate Gamma distribution given in \ref{subsec:stochastic_model}, it is clear that sampling from the latter  requires generating $M$ series of $n$ Gaussian numbers with correlation matrix $\mathbf{G}$. By construction, $\mathbf{G}$ is symmetric and positive definite so that it can be Cholesky factorized in terms of a lower triangular matrix $\mathbf{L}$ such that $\mathbf{G}=\mathbf{L}\mathbf{L}^T$ (where superscript $^T$ denotes matrix transpose) \cite{Devroye}. Once $\mathbf{L}$ is calculated, and if
$Y_1,\ldots, Y_{n}$ are $n$ independent Gaussian numbers, then
$\mathbf{X}=\mathbf{L}\mathbf{Y}$ is a vector of $n$ elements $X_i$ which have
correlation matrix $\mathbf{G}$. This stems from the fact that

\begin{equation}
\mathbf{X}^T\mathbf{G}^{-1}\mathbf{X} = (\mathbf{L}^{-1}\mathbf{X})^T \mathbf{L}^{-1}\mathbf{X}= \mathbf{Y}^T\mathbf{Y},
\end{equation}

\noindent so that by the law of transformation of probability densities, if $\mathbf{Y}$ has a multivariate normal law, $\mathbf{X}=\mathbf{L}\mathbf{Y}$ is distributed according to Eq. (\ref{eq:multivariate_gauss}). We present on Fig. \ref{fig:maps_ex} a 2D map ($100\times 100$ cells) of variables $\nu_i$, obtained from Eq. (\ref{eq:Yi}) using the same sample of $\mathbf{Y}$. The matrix $\mathbf{G}$ (size $10^4\times10^4$) is constructed from Eq. (\ref{eq:G_exp}), with $M=3$ (fluctuation rate $R\simeq 82$ \%) and for values of  $\lambda/\mathcal{L}$ from $10^{-3}$ to 2. The same color scale is used for all four sub-figures. The same underlying pattern can be recognized, but is progressively smeared out as $\lambda$ increases. For values of $\lambda$ large compared to the size of the domain, the field $\nu(\mathbf{r})$ becomes constant in space (but the value taken by $\nu$ is different for each realization).
In practice, straightforward application of this technique in 2D is currently limited to a domain size of the order $\mathcal{N}=10^5$ cells, mainly because of memory issues. In fact, for $n=300$, the correlation matrix has $4\times 10^9$ independent elements, \textit{i.e.} $\sim 30$ Gb in single precision. Cholesky factorization requires $\mathcal{N}^3/3$ operations for a $\mathcal{N}\times \mathcal{N}$ matrix, that is about 100 s for $\mathcal{N}=300\times 300$ on a $80$ GFlop workstation. As a result, we are effectively restricted in 2D to $\lambda/\mathcal{L} \ge 3\times 10^{-2}$. This is clear on Fig. \ref{fig:maps_ex}, where density is constant on a cell of size $\mathcal{L}/100$, whatever the value of $\lambda$. As a result, choosing $\lambda/\mathcal{L}=10^{-3}$ in this case ensures that two neighboring cells are uncorrelated, but the effective correlation length is of order $\mathcal{L}/100$. In 1D, if there are n cells, the correlation matrix is $n\times n$ and $\lambda/\mathcal{L} \ge 10^{-6}$.

\subsection{Source of neutrals}

\begin{figure}[htb]
\includegraphics[width=.4\textwidth]{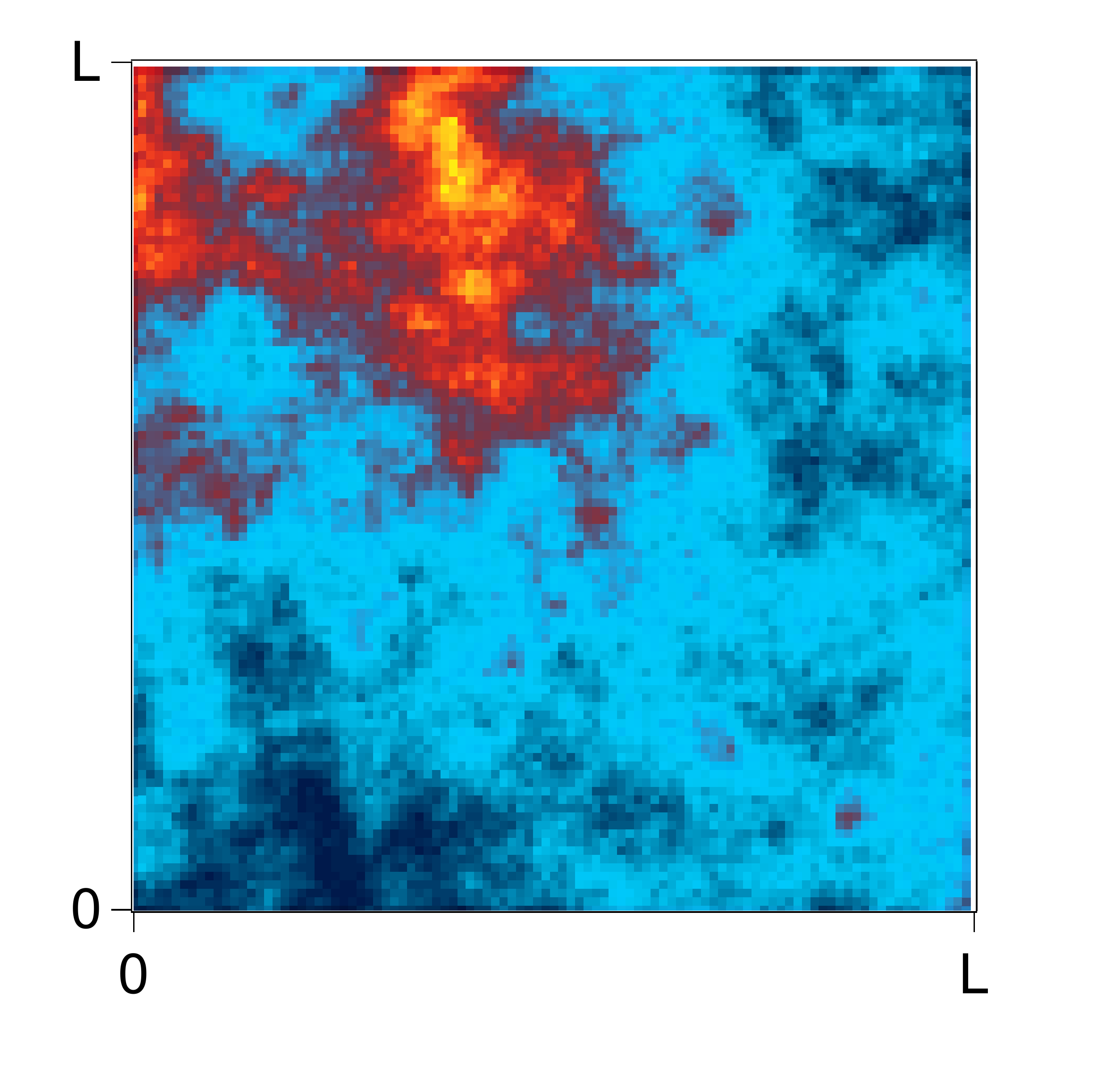}~a)
\includegraphics[width=.45\textwidth]{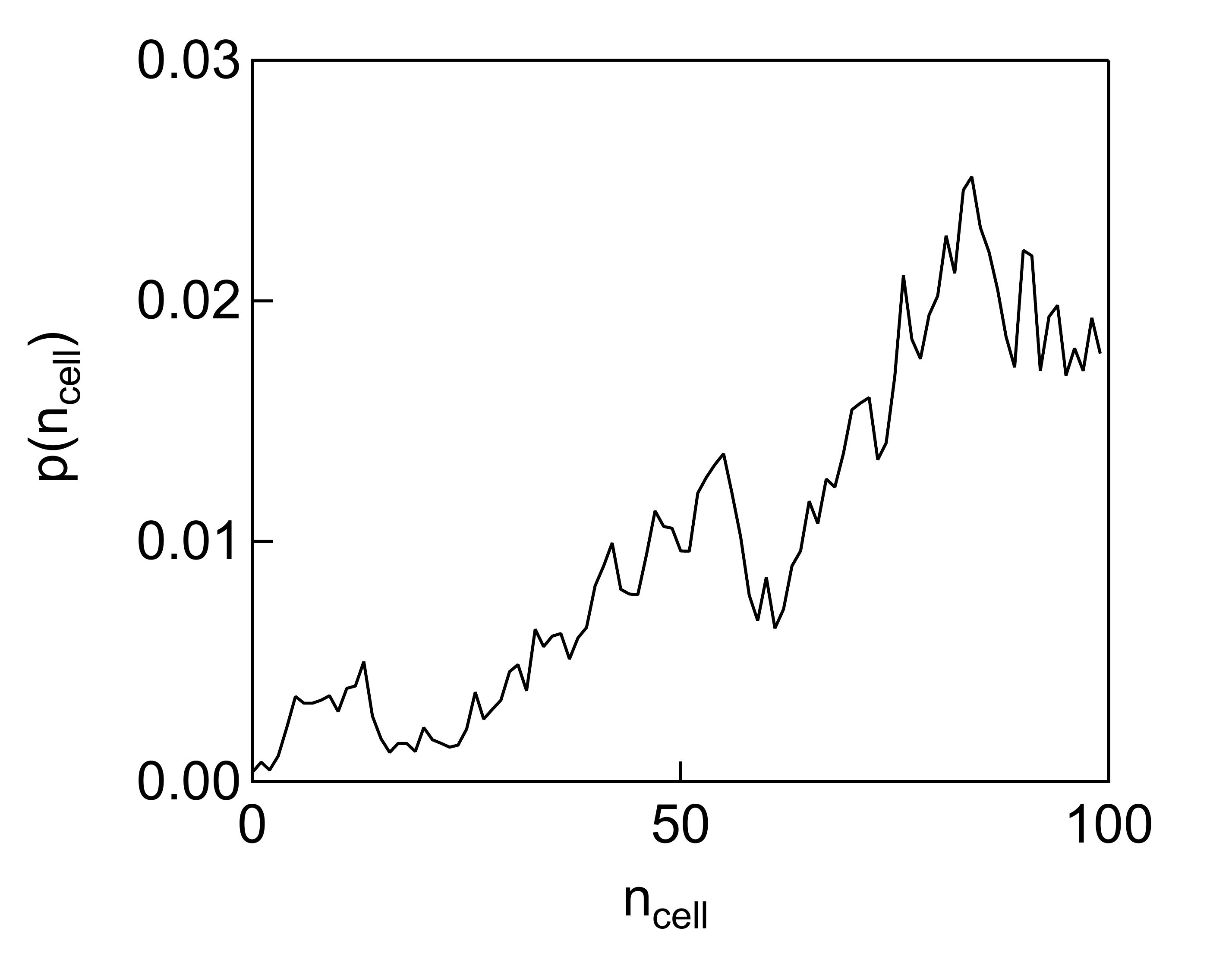}~b)
\caption{a) sampled plasma density map for $\lambda/\mathcal{L}=0.15$, the wall is at $x=0$ b) corresponding neutral birth pdf, plotted as a function of cell number (cell number 100 corresponds to $y=\mathcal{L}$).
\label{fig:recycle}}
\end{figure}

We discuss here the implementation of the stochastic boundary condition introduced in section \ref{sec:boundary}, where the magnitude of the neutral source is proportional to the plasma density at the wall (fast recycling). When the plasma density background has been sampled, $\Gamma_p(y)=N_e(0,y)V_{blob}$ is computed and then used, with the proper normalization, as a PDF for creating neutrals along the $y$ direction, at position $x=0$. The results are finally rescaled using for each realization the total incoming flux by integrating $\Gamma_p$ over y. This procedure is illustrated by plotting a particular realization of the density map on Fig. \ref{fig:recycle} a), and the corresponding PDF on  Fig. \ref{fig:recycle} b).

\subsection{Validation}

The implementation of our stochastic model for the plasma background in the EIRENE Monte Carlo solver  has been checked by running cases corresponding to the model described in the previous sections. Charge exchange and the density dependence of the rate coefficient $\overline{\sigma_{io} v_e}$ are turned off. To simulate a 1D (scattering free) problem, a point source is used in a 2D slab, and all the neutrals are launched with the same velocity $v_0$ along the $x$ axis. In this extremely simplified case,  the conditional expectation estimator implemented in EIRENE  provides a zero variance Monte Carlo scheme, tracking only one particle \cite{Reiter84}. Numerical results for the average density and the standard deviation after many realizations of the stochastic background are plotted on Fig. \ref{fig:av_dens} (open circles) for $a=\infty$ with $M=3$, $6$ and $20$, and for $M=3$ with $a=0.45$, $2.7$, $9$ and $\infty$. In the simulations, neutrals are removed (absorbed) when crossing the $x=\mathcal{L}$ surface. The standard deviations are also identical to those obtained analytically, as shown on Fig. \ref{fig:std_lambda}, again with open circles. The same agreement is obtained for the average ionization source, and the corresponding standard deviation (see Fig. \ref{fig:Sio}). The fast recycling case has been validated against Eq. (\ref{eq:Nav_rec}) and (\ref{eq:Nsq_rec}). Finally, numerical results in 2D are validated against Eq. (\ref{eq:pth_moment_2D}) on Fig. \ref{fig:2Dvs1D}. For these runs, periodic boundary conditions are imposed at $y=0$ and $y=\mathcal{L}$. This exercise provides strong cross checks between the numerical and the analytical results, and allows estimating the number of realizations of the plasma background needed to obtain reliable results for the averages and the standard deviations of the various quantities of interest. In fact, in the worst cases, \textit{i.e.} for large values of $\lambda$, $10^4$ realizations provide very good estimates for the two first moments, but these are still not enough for higher order moments (which can be calculated for the density in the slow recycling case from Eq. (\ref{eq:pth_moment})). In practice, for small values of $\lambda$, less iterations are needed because the density profiles along x are space averages along the y direction, and that different stripes defined by $y_i<y<y_{i+1}$ such that $y_{i+1}-y_i> \lambda$ can in a first approximation be seen as independent realizations.
 
\subsection{Calculations including scattering}

\begin{figure}[htb]
\includegraphics[width=.45\textwidth]{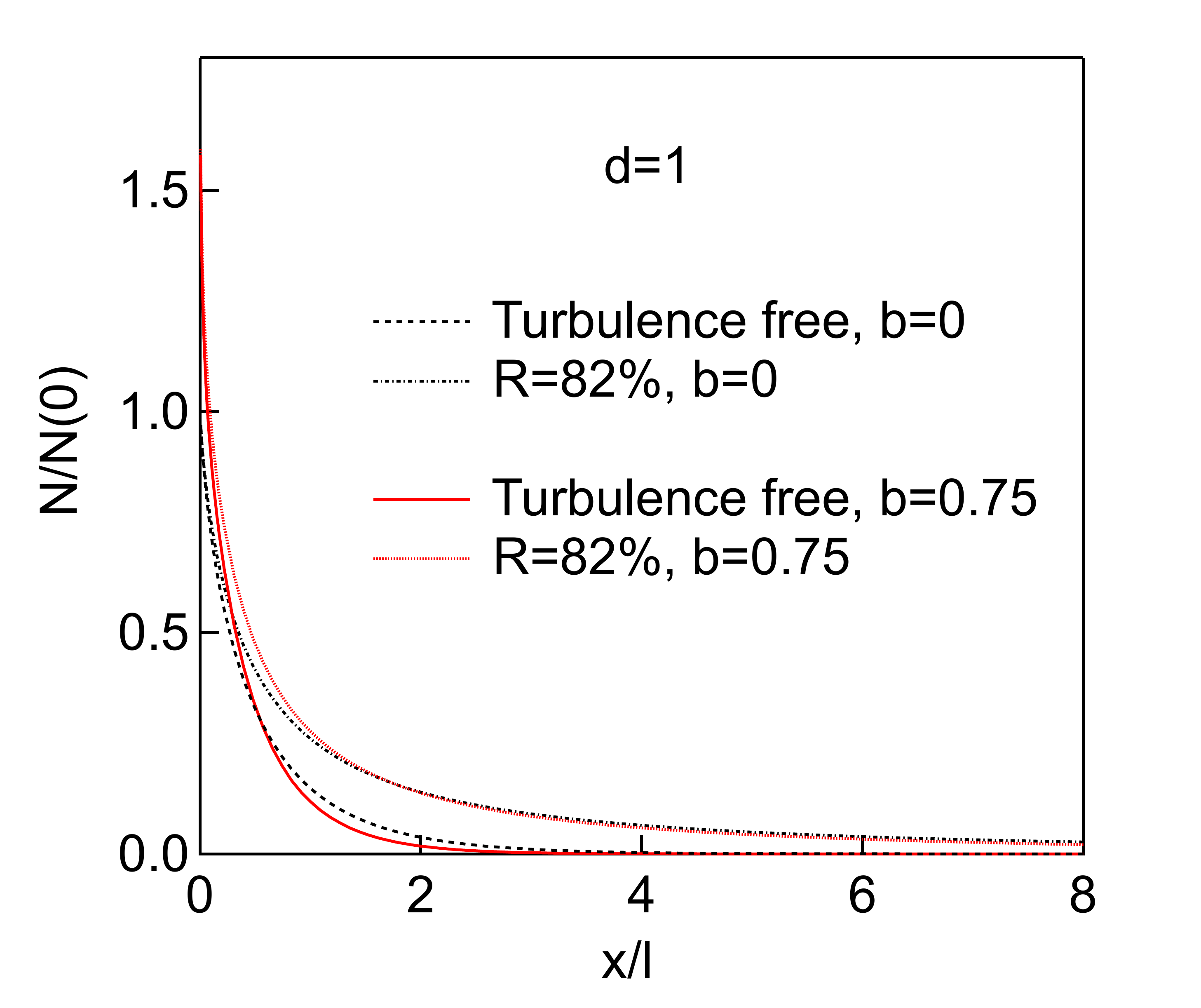}~a)
\includegraphics[width=.45\textwidth]{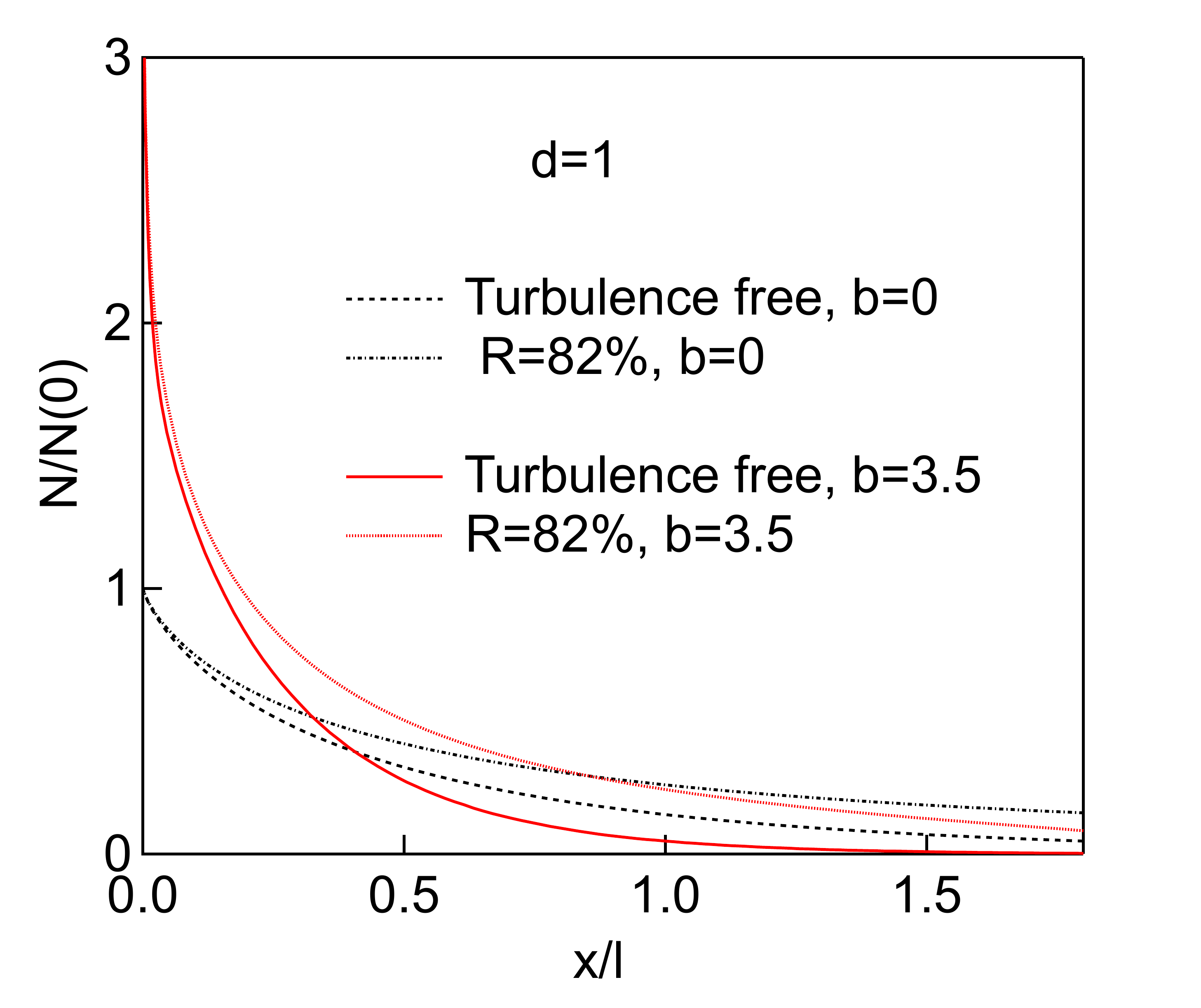}~b)
\caption{a) Density profiles calculated in the slow recycling case for $\lambda=\infty$, $b=0.75$, $d=1$, in the fluctuation free case (solid line) and for $R=82$\% (dotted line), compared to the scattering free results ($b=0$, fluctuation free (dotted line) and $R=82$\% (dash-dotted line)). b) Same as a), but for $b=3.5$.
\label{fig:b_effect_cx}}§
\end{figure}

In this last section, we investigate whether taking scattering into account changes the overall picture that was obtained in the 2D scattering free case. We limit ourselves to the effect of density fluctuations. In the case of D atoms in a plasma, this amounts to neglect both ion temperature and velocity fluctuations, since the properties of charge exchange (CX) neutrals depend on those of the background ion velocity distribution. In the edge plasmas of tokamaks, a simple estimate of the turbulent velocities shows that the latter are small compared to the thermal velocity \cite{Marandet06}, so that the velocity of CX neutrals should not be strongly affected by turbulence. In an homogeneous background medium, if scattering conserves kinetic energy (one speed transport model), the neutral decay length should be smaller than in the scattering free case because trajectories are no longer ballistic. In the limit of vanishing scattering mean free path (compared to plasma inhomogeneities), the transport becomes diffusive. If the initial neutral velocity $v_0$ differs from the thermal ion velocity $v_{th}$, neutrals which have undergone charge exchange have different mean free paths than first generation neutrals. The latter have a  parameter $a$ calculated from $E_0=1/2 mv_0^2$, while all subsequent generations have a value $a$ calculated from $3T_i/2$. To summarize, when scattering is retained, at least two more parameters enter the problem. The first one is the ratio of the scattering to the absorption rate, $b=\overline{\sigma_{s}v}/\overline{\sigma_{a}v}$, where $\sigma_a$ and $\sigma_s$ are the corresponding cross sections. The second parameter, denoted by $d$, is the ratio of the energy of scattered neutrals to their initial energy, namely $d=3T_i / 2E_0$ (in the one speed problem, the velocity after charge exchange is calculated from $v_{cx}=\sqrt{3kT_i/m}$ so as to ensure proper thermalization). It should be noted that $b$ and $d$ are closely related to the parameters which control the analytical solution of Boltzmann's equation obtained in simplified 1D geometry with mono-kinetic ions by Smirnov, namely $\beta=\overline{\sigma_{s}v}/(\overline{\sigma_{a}v}+\overline{\sigma_{s}v})$ and $u_0$, the neutral initial velocity in units of ion thermal velocity \cite{Rehker73}. In the calculations presented below, periodic boundary conditions are implemented for $y=0$ and $y=\mathcal{L}$, the $x=0$ surface is reflecting and $x=\mathcal{L}$ absorbing.   The effect of $b$ is investigated on figures \ref{fig:b_effect_cx}~a) and b) for $d=1$, where results respectively obtained for $b=0.75$ and $b=3.5$ are compared to those obtained for the scattering free case ($b=0$). 
\begin{figure}[htb]
\includegraphics[width=.45\textwidth]{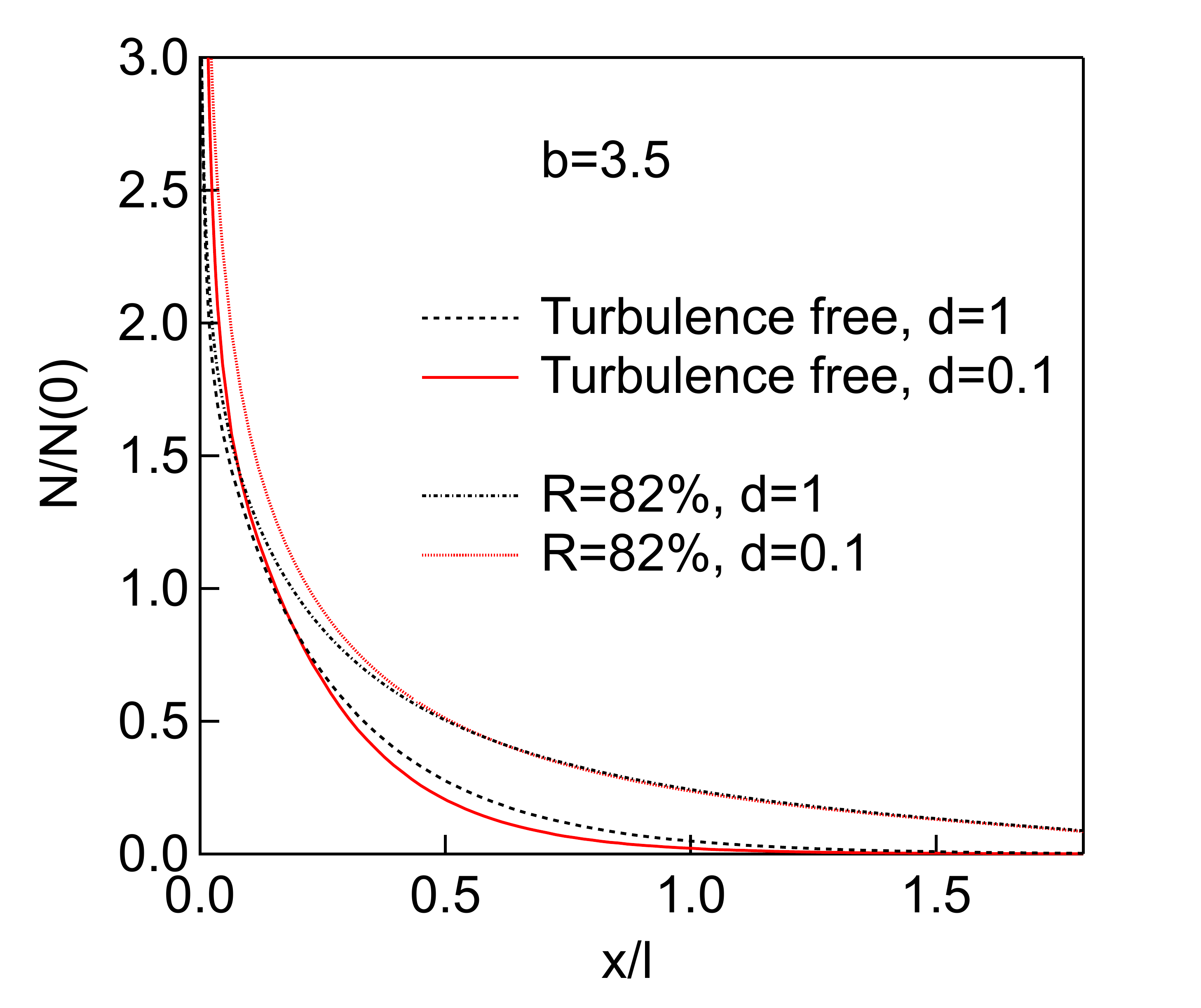}~a)
\includegraphics[width=.45\textwidth]{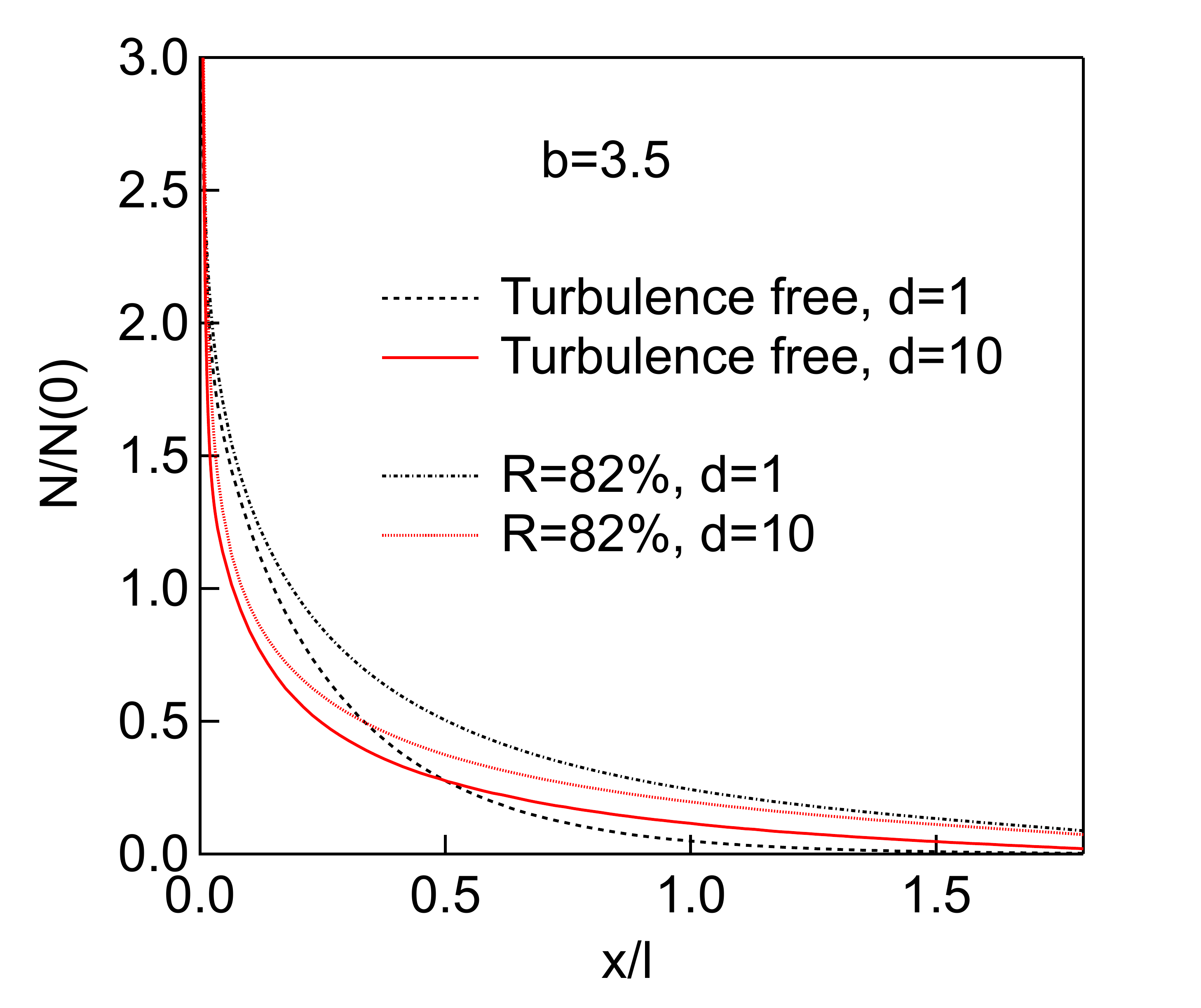}~b)
\caption{a) Density profiles calculated in the slow recycling case for $\lambda=\infty$, $b=3.5$, $d=0.1$, in the fluctuation free case (solid line) and for $R=82$\% (dotted line), compared to the $d=1$ case, fluctuation free (dotted line) and $R=82$\% (dash-dotted line). b) Same as a), but for $d=10$.
\label{fig:d_effect_cx}}
\end{figure}
The calculations confirm that the neutral density is larger close to the wall when scattering is included, and show that the effects of fluctuations are similar than in the scattering free case, even for $b=3.5$. The effect of the parameter $d$ is shown on Fig. \ref{fig:d_effect_cx}, where density profiles are plotted for $b=3.5$, $d=0.1$ and $d=10$, and compared to the $d=1$ case (for which the neutral particle energy does not change in the scattering event). In the latter case, the effect of fluctuations is clearly reduced compared to $d=1$ or $d=0.1$, as expected. In fact, after charge exchange, the neutral particle have a higher energy, hence the ratio $a$ of the turbulent integral scale to the mean free path is smaller.

\section{Conclusions and perspectives}

We have presented a stochastic model which allows investigating linear transport in a turbulent background. The statistical properties of turbulence have been described by a multivariate Gamma law, because the latter has three major attractive features. First, averages can be calculated analytically in the scattering free case, which is relevant for molecules and sputtered neutral impurities in tokamaks.  Next, it is conveniently implemented in a Monte Carlo code, here the EIRENE Monte Carlo Solver, so that analytical and numerical results can be cross-checked. Finally, the univariate Gamma distribution (\textit{i.e.} the one point marginal) provides a good description of plasma density fluctuations in the outer edge of tokamaks. The multivariate Gamma law is fully specified by its correlation matrix, which is closely related to the spatial spectra of fluctuations. In the frame of this stochastic model, the average neutral density and ionization source profiles have been investigated. In the scattering free case, two main control parameters have been identified. The first one is related to the properties of the source, namely whether it is stochastic or not. In the frame of neutral transport in plasmas, this is controlled by the ordering between recycling and turbulence time scales. We thus distinguish two limiting cases, namely slow and fast recycling, the relevance of which depends both on the elementary process at play and the wall status (\textit{i.e.} saturated in hydrogen or not). The second control parameter is the ratio $a$ of the turbulence correlation length (or integral scale) to the average neutral mean free path. For small values of $a$, the ergodic theorem implies that the average quantities can be calculated by replacing the absorption rate by its average in Boltzmann's equation, thus recovering the so-called atomic-mix limit in Ref. \cite{Pomraning91}. This does not mean that turbulence has no effect, since the average ionization rate can differ from the ionization rate calculated for the average plasma parameters, in particular when temperature fluctuations are present. For non-stochastic boundary conditions, fluctuations enhance neutral penetration, in the sense that the decay of the average density profile is slower than in the turbulence free case. This effect becomes more and more pronounced  as $a$ is increased. The differences between the average ionization source, and the ionization source calculated in the turbulent free case are less significant. The physical reason for this is clear. Realizations which have typically low density lead to deep neutral penetration, but at the same time to low values for the ionization rate. In other words, the source, being homogeneous in space, is significant even in regions where the plasma has low density. In contrast, in the fast recycling case the source is concentrated where the density is high, so that neutrals are more likely to be ionized close to the wall provided the electron temperature is high enough. In this case, the ionization source becomes more peaked towards the wall. These conclusions remain valid in the presence of scattering, as shown by numerical calculations. They provide a clear physical picture of the role of fluctuations on neutral transport, and are therefore expected to remain qualitatively correct when fluctuations obey a different statistics. The effect of turbulence is significant only for large fluctuation rates ($> $ 50 \%). In the fusion context, this means that our results are of interest mainly for SOL plasmas. The confined plasma might nevertheless be affected, through a reduction of the SOL screening efficiency. \\

The work presented here is a first step towards evaluating the importance of turbulence on neutral particle transport in fusion plasmas. Further work will now focus on implementing a more realistic description of the outer edge of the plasma, \textit{i.e.} to relax both the statistical homogeneity and stationarity assumption. The role of temperature fluctuations must also be thoroughly studied, because of threshold effects in the ionization cross section. Both can be addressed in the frame of our stochastic model, or rely on the output of a turbulence code. A direct coupling to such a code would finally allow to relax the passive assumption, \textit{i.e.} to investigate the back reaction of neutrals on turbulence.

\begin{acknowledgments}
This work is supported by the Agence Nationale de la Recherche (ANR-07-BLAN-0187-01, project PHOTONITER), and  is part of a collaboration (LRC DSM99-14) between the laboratoire de Physique des Interactions Ioniques et Mol\'{e}culaires PIIM (UMR 6633) and the Institut de Recherche Fusion par Confinement Magn\'{e}tique - ITER, CEA Cadarache, within the framework of the F\'{e}d\'{e}ration de Recherche sur la Fusion Magn\'{e}tique (FR-FCM). J. Rosato is supported through an EFDA Fellowship, under task agreement WP08-FRF-CEA-FZJ-Rosato.
\end{acknowledgments}

\appendix

\section{Expression of the chi-2 PDF}

We consider here the one point marginal $W_1$ of $W_N$ defined by Eq. (\ref{eq:multi_chi-2}). The integration of $W_N$ over $n-1$ variables is straightforward, and leads to

\begin{equation}\label{eq:def_chi2}
W_1(\nu)=\int dX_1\ldots \int dX_M \frac{1}{(\sqrt{2\pi}\sigma)^M}\exp\left(-\frac{\sum_{i=1}^M X_i^2}{2\sigma^2}\right) \delta\left(\nu-\sum_{i=1}^M X_i^2\right).
\end{equation}

\noindent To calculate $W(\nu)$ we note that the integral is of the form

\begin{equation}
W_1(\nu)=\int dX_1\ldots \int dX_M P(X_1,\ldots,X_M) \delta\left(f(X_1,\ldots,X_M)\right).
\end{equation}

\noindent Using the co-area formula, we get

\begin{equation}
W_1(\nu)=\int_U dS(\mathbf{X}) \frac{1}{|\bigtriangledown f|}P(X_1,\ldots,X_M),
\end{equation}

\noindent where $U$ is the set of point in $R^M$ such that $f(\mathbf{X})=0$, and $S$ a measure on this set. Now, all the terms in the integral involve only $\sum_{i=1}^M X_i^2$ and can simply be taken out, hence 

\begin{equation}
W_1(\nu)=\frac{1}{(\sqrt{2\pi}\sigma)^M}\frac{1}{2\nu^{1/2}}\exp\left(-\frac{\nu}{2\sigma^2}\right)\int_U dS(\mathbf{X}).
\end{equation}

\noindent The remaining integral is the area of the hypersphere of radius $r=\nu^{1/2}$ in dimension $M$, so that

\begin{equation}
W_1(\nu)=\frac{1}{(2\sigma^2)^{M/2} \Gamma\left(\frac{M}{2}\right)}\nu^{M/2-1}\exp\left(-\frac{\nu}{2\sigma^2}\right).
\end{equation}

\noindent which is the usual expression of the chi-2 distribution with M degrees of freedom and scale parameter $\sigma$ (\cite{Kotz_uni}, p. 452). The later is nothing else but a Gamma distribution with a shape factor $\alpha=M/2$, a scale factor $2\sigma^2$ and a displacement $\gamma=0$ (\cite{Kotz_uni}, p. 337).

\section{Limiting cases for the average density profiles}

We investigate here three limiting cases of Eq. (\ref{eq:pth_moment}) with $p=1$ and of Eq. (\ref{eq:Nav_rec}), namely the small fluctuation limit, the infinite correlation length limit ($\lambda\rightarrow\infty$), and its opposite $\lambda\rightarrow 0$. First, the vanishing fluctuations limit corresponds to $M\rightarrow +\infty$ and $G_{ij}\propto 1/M \rightarrow 0$ (\textit{i.e.} $\sigma = \nu_0\sqrt{2/M}$), so that $\langle \nu\rangle\rightarrow \nu_0$ (where $\nu_0$ is the ionization rate in the turbulence free case) and $\langle\langle\nu^2\rangle\rangle\rightarrow 0$. As $M$ becomes large, 

\begin{equation}
\det(\mathbf{I}+2\mathbf{G} \mathbf{U}_k)\simeq 1+2Tr(\mathbf{G} \mathbf{U}_k)=1+\frac{2\epsilon k \nu_0}{M v_0},
\end{equation}

\noindent so that using $x_k=k\epsilon$ Eq. (\ref{eq:pth_moment}) reduces to

\begin{equation}\label{eq:vanish_fluc}
   \langle N(x_k)\rangle \simeq \frac{\Gamma_0}{v_0}\exp - \nu_0 x_k /v_0,
\end{equation}

\noindent as it should. In the case of Eq. (\ref{eq:Nav_rec}), the limit of $\varpi$ must also be studied. For small fluctuations,  we have $\mathbf{A}^{-1}\simeq \mathbf{I}-2\mathbf{G}\mathbf{U}_k$, hence  $1-(A^{-1})_{ll}\simeq 2\epsilon\langle\nu\rangle /Mv_0\delta_{l\leq k}$, so that $\varpi$ reduces to $\langle\nu\rangle$. Now we consider the $\lambda\rightarrow 0$ limit, where $\mathbf{G}$ tends to a diagonal matrix $\mathbf{G}_d$ such that $(\mathbf{G}_d)_{ii}=\sigma/\sqrt{2M}$. The calculation of the determinant is then trivial, and we obtain

\begin{equation}
\langle N_0(x_k)\rangle=\frac{\Gamma_0}{v_0}\left[\frac{1}{(1+2 x_k G_0/ k v_0 )^k}\right]^{M/2},
\end{equation}

\noindent where $G_0=\langle \nu\rangle$/M. But when $\lambda\rightarrow 0$, we must also impose $\epsilon\rightarrow 0$, $k\rightarrow\infty$, $k\epsilon=x$, so that the average density profile reduces to

\begin{equation}\label{eq:vanish_fluc}
   \langle N_0(x)\rangle \simeq \frac{\Gamma_0}{v_0}\exp - \langle\nu\rangle x /v_0,
\end{equation}

\noindent as expected from the ergodic theorem. In the the fast recycling case, it is straightforward to show that $\varpi\rightarrow\langle \nu\rangle$, so that Eq. (\ref{eq:Nav_rec}) also reduces to Eq. (\ref{eq:vanish_fluc}). Finally, we study the infinite correlation length limit $\lambda\rightarrow \infty$. In this limit, $\mathbf{G}\mathbf{U}_k$ becomes proportional to a $n\times n$ matrix $\mathbf{B}_k$ such that its elements are equal to $(B_k)_{ij}=\delta_{j \le k}$. This matrix has two eigenvalues, namely 0 with rank n-1 and k with rank 1. The calculation of the determinant is then again trivial, and leads to

\begin{equation}\label{eq:N_av_lambda_inf}
\langle N_{\infty}(x_k)\rangle=\frac{\Gamma_0}{v_0}\left[\frac{1}{1+2 x_k G_0 / v_0 }\right]^{M/2}.
\end{equation}

\noindent For large correlation lengths, the plasma background becomes uniform, so that this result can be recovered from a simple integral, namely

\begin{equation}
\langle N_{\infty}(x)\rangle=\frac{\Gamma_0}{v_0}\int_0^{+\infty} d\nu W_1(\nu) \exp(-\nu x/v_0),
\end{equation}

\noindent where $W_1$ is the univariate chi-2 distribution with M degrees of freedom given by Eq. (\ref{eq:def_chi2}). Furthermore, for $\lambda\rightarrow +\infty$, $\mathbf{A}_k$ can be expressed as $\mathbf{A}_k=I+(2G_0\epsilon/v_0) \mathbf{B}_k$. It is easily shown by induction that $(\mathbf{B}_k)^n=k^{n-1}\mathbf{B}_k$, so that  expanding $\mathbf{A}_k ^{-1}$ in Neumann series leads to

\begin{equation}
\mathbf{A}_k^{-1}=\sum_{n=0}^{+\infty}(-1)^n \left(\frac{2\epsilon G_0}{v_0}\right)^n (\mathbf{B}_k)^n=I-\frac{2\epsilon G_0}{v_0}\left(1+\frac{2 G_0 x_k}{v_0}\right)^{-1} \mathbf{B}_k,
\end{equation}

\noindent where $x_k=\epsilon k$. In the continuum limit where $\epsilon\rightarrow 0$, $k\rightarrow +\infty$ and $x_k=x$, the effective ionization rate $\varpi$ reduces to

\begin{equation}\label{eq:prefac_lambda_inf}
\varpi=\langle\nu\rangle \left(1+\frac{2 G_0 x_k}{v_0}\right)^{-1}.
\end{equation}
 
\noindent This result is consistent with that obtained from the integral

\begin{equation}
\langle N_{\infty}^{rec}(x)\rangle=\frac{\Gamma_0}{v_0\langle\nu\rangle}\int_0^{+\infty} d\nu W_1(\nu) \nu\exp(-\nu x/v_0).
\end{equation}
 
\section{Asymptotic behavior of $\varpi(x)$, an application of Szeg\"{o}'s theorem}\label{app:Szego}

 Consider a set of symmetric Toeplitz matrices $T_n$ ($n\times n$) with $n=1,\ldots,+\infty$, \textit{i.e.} matrices such that $(T_n)_{ij}=(T_n)_{i-1,j-1}$ for $i>2$. A Toeplitz matrix is completely specified by $2n-1$ entries $t_m$, with $m\in[-n-1,n+1]$, where positive values of m refer to the upper part of the matrix. For a symmetric matrix $t_{-m}=t_m$. We define a function $f(p)$ by

\begin{equation}\label{eq:fp}
f(p)=\sum_{m=-\infty}^{+\infty} t_m e^{ipm},
\end{equation}

\noindent with $p\in [0,2\pi]$. Szeg\"{o}'s theorem implies \cite{Gray}

\begin{equation}\label{eq:det_ratio}
\lim_{n\rightarrow +\infty}\frac{\det T_{n-1}}{\det T_n} =  \exp -\frac{1}{2\pi} \int_0^{2\pi} dp \ln f(p)=L_{\infty}.
\end{equation}

\noindent We apply this result to a set of $k\times k$ Toeplitz matrices $T_k$ such that $T_k=I_k+2\epsilon/v_0 G_k$. In this case

\begin{equation}\label{eq:t_m}
t_m=\delta_{m0}+G_0\exp-\frac{m \epsilon}{2\lambda},
\end{equation}

\noindent where $G_0=\sigma/\sqrt{2M}$. It is easily checked that $\det(T_k)=\det(A_k)$ (where we recall that $A_k=I_n+2GU_k$ is a $n\times n$ matrix ). The matrix elements $(A_k^{-1})_{ii}$  can be written in terms of the corresponding cofactors $\mathcal{C}_{ii}$, as

\begin{equation}
(A_k^{-1})_{ii}=(-1)^{2i} \frac{\mathcal{C}_{ii}}{\det(T_k)}=\frac{\det(T_{k-1})}{\det(T_{k})},
\end{equation}

\noindent where the last equality holds for $i=1$ and $i=k$ only. Therefore, for large $k$, $(A_k^{-1})_{11}\sim L_{\infty}$. Furthermore, $(A_k^{-1})_{11}=(A_k^{-1})_{kk}$ because the corresponding minors, obtained by removing either the first or the last line and column of the $A_k$ matrix, are equal. A straightforward application of Eq. (\ref{eq:fp}) with the $t_m$ defined by Eq. (\ref{eq:t_m}) leads to

\begin{equation}
f(p)=1+\frac{2\epsilon}{v_0}G_0 \tanh\left(\frac{\epsilon}{2\lambda}\right)\frac{1}{1-\cos(p)/\cosh(\epsilon/2\lambda)}.
\end{equation}

\noindent The integral in Eq. (\ref{eq:det_ratio}) can be calculated analytically, yielding

\begin{equation}\label{eq:discrete_Szego}
L_{\infty}=\frac{(1+\tanh(\epsilon/2\lambda))}{ g+\left(g^2-1/\cosh^2 (\epsilon/2\lambda)\right)^{1/2}},
\end{equation}
 
\noindent where
 
\begin{equation}
g=1+\frac{2\epsilon}{v_0}G_0 \tanh\left(\frac{\epsilon}{2\lambda}\right).
\end{equation}

\noindent Finally, in the continuous limit where $\epsilon\rightarrow 0$ (in practice $\epsilon/\lambda\ll 1$),

\begin{equation}
\lim_{\epsilon\rightarrow 0} \frac{1- L_{\infty}}{\epsilon} = \frac{1}{2\lambda}\left[\left(1+\frac{8G_0\lambda}{v_0}\right)^{1/2}-1\right],
\end{equation}

\noindent from which Eq. (\ref{Eq:Szego_continuous}) is obtained.

\end{document}